\documentclass[aip,preprint,amsmath,amssymb]{revtex4-2} %%,footinbib
\usepackage{graphicx}% Include figure files
\graphicspath{{figures/}}
\usepackage{dcolumn}% Align table columns on decimal point
\usepackage{bm}% bold math
\usepackage{amsmath}
\usepackage{amssymb}
\usepackage{txfonts}
\usepackage{url}
\usepackage{braket}
\usepackage{multirow}
\usepackage[normalem]{ulem}
\usepackage{here}
\usepackage[usenames,dvipsnames]{color}
\definecolor{Gray}{gray}{0.0}
\definecolor{lightGray}{gray}{0.35}
\usepackage[linesnumbered,ruled,vlined]{algorithm2e}

\makeatletter
\def\Hline{%
\noalign{\ifnum0=`}\fi\hrule \@height 1pt \futurelet
\reserved@a\@xhline}
\makeatother

\begin{document}
\title{Load-Balanced Diffusion Monte Carlo Method with Lattice Regularization}
%
% Authors
%
\author{Kousuke Nakano}
\email{kousuke\_1123@icloud.com}
\affiliation{Center for Basic Research on Materials (CBRM), National Institute for Materials Science (NIMS), 1-2-1 Sengen, Tsukuba, Ibaraki 305-0047, Japan}
\author{Sandro Sorella}
\affiliation{International School for Advanced Studies (SISSA), Via Bonomea 265, 34136, Trieste, Italy}
\author{Michele Casula}
\affiliation{Institut de Min{\'e}ralogie, de Physique des Mat{\'e}riaux et de Cosmochimie (IMPMC), Sorbonne Universit{\'e}, CNRS UMR 7590, IRD UMR 206, MNHN, 4 Place Jussieu, 75252 Paris, France}

%%%%%%%%%%%%%%%%%%%%%%%%%%%%%%%%%%%%%%%%%%%%%%%%%%%%%%

\date{\today}
\begin{abstract}
\emph{Ab initio} quantum Monte Carlo (QMC) is a stochastic approach for solving the many-body Schrödinger equation without resorting to one-body approximations. QMC algorithms are readily parallelizable via ensembles of $N_w$ walkers, making them well suited to large-scale high-performance computing.
Among the QMC techniques, Diffusion Monte Carlo (DMC) is widely regarded as the most reliable, since it provides the projection onto the ground state of a given Hamiltonian under the fixed-node approximation. One practical realization of DMC is the Lattice Regularized Diffusion Monte Carlo (LRDMC) method, which discretizes the Hamiltonian within the Green’s Function Monte Carlo framework.
DMC methods — including LRDMC — employ the so-called branching technique to stabilize walker weights and populations. At the branching step, walkers must be synchronized globally; any imbalance in per-walker workload can leave CPU or GPU cores idle, thereby degrading overall hardware utilization. The conventional LRDMC algorithm intrinsically suffers from such load imbalance, which grows as $\log(N_w)$, rendering it less efficient on modern parallel architectures.
In this work, we present an LRDMC algorithm that inherently addresses the load imbalance issue and achieves significantly improved weak-scaling parallel efficiency.
Using the binding energy calculation of a water–methane complex as a test case, we demonstrated that the conventional and load-balanced LRDMC algorithms yield consistent results. Furthermore, by utilizing the Leonardo supercomputer equipped with NVIDIA A100 GPUs, we demonstrated that the load-balanced LRDMC algorithm can maintain extremely high parallel efficiency ($\sim$98\%) up to 512 GPUs (corresponding to $N_{\rm w}= 51200$), together with a speedup of $\times~1.24$ if directly compared with the conventional LRDMC algorithm with the same number of walkers. The speedup stays sizable, i.e., $\times~1.18$, even if the number of walkers is reduced to $N_{\rm w}=400$.
\end{abstract}
\maketitle

%%%%%%%%%%%%%%%%%%%%%%%%%%%%%%%%%%%%%%%%%%%%%%%%%%%%%%%%%%%
%
% Introduction
%
%%%%%%%%%%%%%%%%%%%%%%%%%%%%%%%%%%%%%%%%%%%%%%%%%%%%%%%%%%%

\section{Introduction}

% Very general introduction of ab initio electronic structure calculations
Understanding and predicting the quantum behavior of electrons and nuclei remains a fundamental challenge in physics and chemistry. Despite decades of theoretical and computational developments, solving the many-body Schrödinger equation exactly is still intractable for systems of practical interest due to its exponential complexity. As a result, approximate methods have become indispensable. Among them, Density Functional Theory (DFT)\cite{1965KOH} has emerged as a standard approach by mapping the complex many-electron problem to a system of non-interacting electrons subject to an effective potential, derived from the so-called eXchange-Correlation functional (XC)~\cite{2004MAR}. However, the accuracy of DFT hinges on the choice of XC functionals, for which no exact form is known. Thus, DFT often struggles in systems with strong electron correlation or non-local interactions, and systematic improvements remain elusive~\cite{2001PER,2017MED}.

\vspace{1mm}
Ab initio Quantum Monte Carlo (QMC) methods aim at solving the many-body Schrödinger equation using stochastic techniques, without relying on one-body approximations. Their intrinsic scalability and accuracy have always been appealing since their first appearance. Among various QMC implementations, Diffusion Monte Carlo (DMC) is particularly notable for its ability to compute the exact ground state energy for a Hamiltonian with boundaries usually defined by the nodal surface of a trial wavefunction, via imaginary-time projection~{\cite{2001FOU}}. This constraint is known as ``fixed-node approximation'' (FNA), 
a remedy for the Fermionic sign problem appearing in the Hamiltonian quantum evolution~{\cite{1982REY}}. These boundaries are in principle systematically improvable, according to the quality of the trial wavefunction nodes, thanks to the FNA variational property. Owing to its quantitative reliability, DMC has been widely applied to challenging systems where DFT often fails, such as solid and liquid hydrogen under high pressure~{\cite{2015DRU, 2023NIU, 2023LOR, 2024TEN}}, molecular crystals~{\cite{2018ZEN, 2024PIA}}, and two-dimensional materials~{\cite{2015MOS, 2022NIK}}.
The most commonly used DMC implementation employs the Suzuki–Trotter decomposition to iteratively perform projection steps~{\cite{2001FOU}}. However, it suffers from numerical instability in the extrapolation to the zero time-step limit, where the total energy does not converge monotonically. To address this issue, an alternative approach based on a lattice regularized Hamiltonian, and thus dubbed Lattice Regularized  Diffusion Monte Carlo (LRDMC), was proposed~{\cite{2005CAS}}. LRDMC exploits the Green's Function Monte Carlo (GFMC) technique{~\cite{1995TEN, 1998BUO, 2000SOR}}—traditionally used in lattice models—to continuous-space systems by discretizing the Laplacian on a real-space mesh.
In LRDMC, the projection operator is implemented as powers of a Green's function, and unlike standard DMC, it does not require Suzuki–Trotter factorization. This eliminates the time-step bias inherent in the conventional DMC implementation. Instead, a bias arises due to the lattice discretization of the Hamiltonian, but this is of the order of $\mathcal{O}(a^2)$ and can be systematically removed by extrapolation to the continuous space limit ($a \rightarrow 0$), which turns out to be much smoother than the time-step extrapolation of standard DMC~\cite{2025PIA}.

{\vspace{1mm}}
In both DMC and LRDMC, the so-called branching step plays a central role in enhancing computational efficiency. Each walker carries out {\emph{independent}} projection steps and accumulates a statistical weight, which is used to perform weight and population control through branching: walkers with large weights are spawned, while those with small weights are removed at periodic intervals. Except for the branching steps, all walkers evolve independently, making these methods highly parallelizable. In practice, large-scale DMC simulations typically involve a large number of walkers distributed across many CPUs or GPUs.
However, branching steps require synchronization: all walkers must complete their projection steps before branching can proceed. If computational load becomes imbalanced across walkers—for example, due to the stochastic nature of the projection—the faster walkers must wait for the slower ones, causing some working units (on CPUs or GPUs) to remain idle. As detailed in the Method section, the conventional LRDMC algorithm inherently leads to this kind of load imbalance, especially as the number of walkers increases. Consequently, its parallel efficiency degrades on modern hardware platforms that rely on a large walker population.
In this work, we address this scalability issue by developing and presenting a new LRDMC algorithm that eliminates load imbalance, thereby enabling efficient use of large number of walkers on modern parallel architectures.

\vspace{1mm}
This paper is organized as follows: in Sec.{{~\ref{methods}}}, we review the conventional LRDMC algorithm and we introduce the new one satisfying load-balance among walkers;
in Sec.{~\ref{validation}}, we verify that the proposed load-balanced LRDMC algorithm gives consistent DMC energies if compared with the conventional LRDMC algorithm for the methane-water dimer.
In Sec.{~\ref{weak-scaling}}, we show how the load balance fulfilled by the new algorithm improves
the weak-scaling properties of the LRDMC algorithm.
In Sec.{~\ref{conclusion}}, we summarize the main outcome of this work. 
%

%%%%%%%%%%%%%%%%%%%%%%%%%%%%%%%%%%%%%%%%%%%%%%%%%%%%%%%%%%%
%
%  Methods
%
%%%%%%%%%%%%%%%%%%%%%%%%%%%%%%%%%%%%%%%%%%%%%%%%%%%%%%%%%%%

\section{Methods}
\label{methods}

\begin{figure*}
    \centering
    \includegraphics[width=1.0\linewidth]{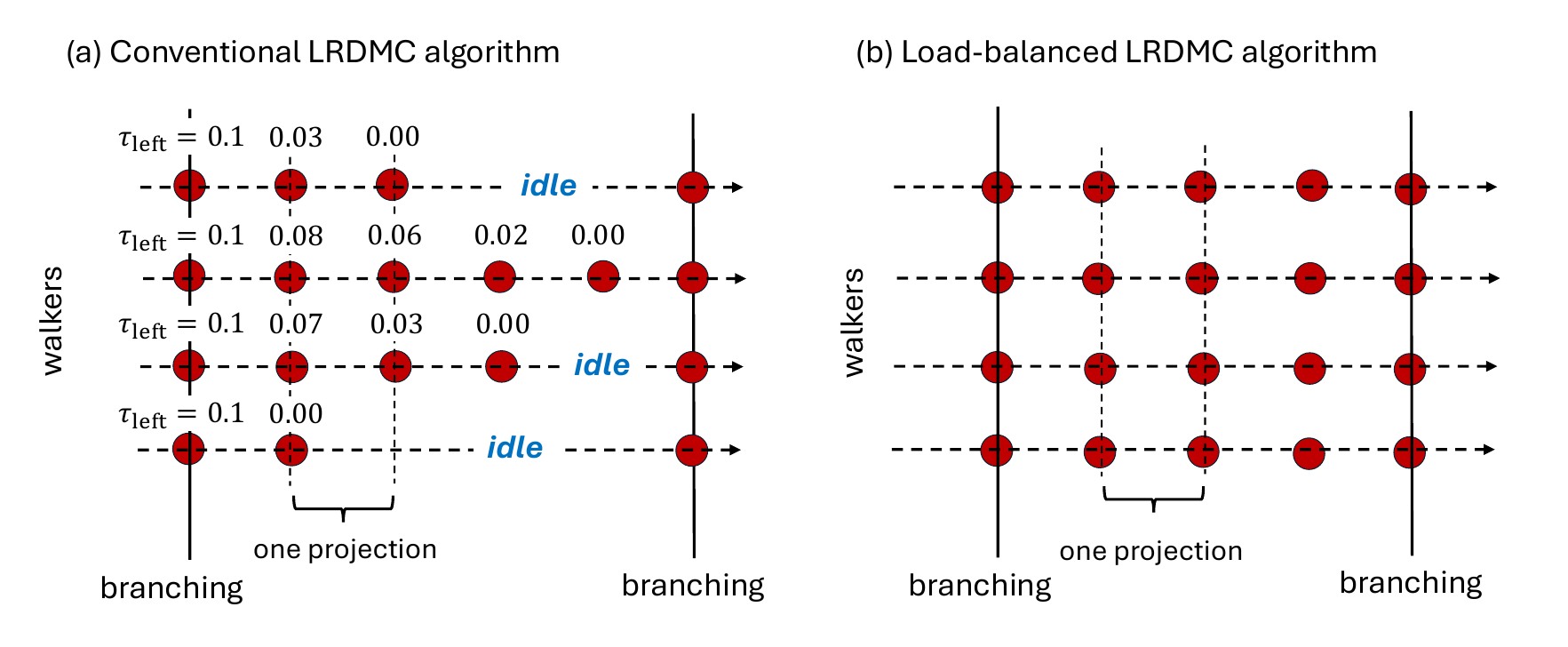}
    \caption[]{The schematic figure of the difference between two LRDMC algorithms. (a) conventional and (b) load-balanced LRDMC algorithms. In algorithm (a), each walker maintains its own remaining time $\tau_{\rm left}$ and undergoes projection steps drawn from independent random numbers until $\tau_{\rm left} = 0.0$, which can leave some walkers idle. In contrast, in algorithm (b) every walker performs the same number of projection steps, thereby preserving load balancing.}
    \label{fig:schematic-lrdmc-algorithms}
\end{figure*}

In this Section, we outline the conventional LRDMC algorithm that has been used in previous studies~{\cite{2019NAK, 2020NAK_LRDMC, 2023RAG, 2024NAK, 2025NAK}}, as well as the load-balanced LRDMC algorithm proposed in this work. The essential difference between the two algorithms lies in whether load balance among walkers is inherently guaranteed when processed in parallel. A schematic comparison of the two algorithms is provided in Fig.~{\ref{fig:schematic-lrdmc-algorithms}}. The mathematical background of both algorithms and the details of their convergence properties are presented in the Appendices \ref{app:stationary_distribution_standard} and \ref{app:stationary_distribution_load_balance}.

%%%%%%%%%%%%%%%%%%%%%%%%%%%%%%%%%%%%%%
% Figure
%%%%%%%%%%%%%%%%%%%%%%%%%%%%%%%%%%%%%%

\subsection{Overview of the Lattice regularized diffusion Monte Carlo method}
\label{subsec:gfmc-and-lrdmc}
Lattice regularized diffusion Monte Carlo (LRDMC)~{\cite{2005CAS}} is a projection technique that allows one to improve a variational ansatz systematically in the continuous space. This method is based on Green's function Monte Carlo (GFMC){~\cite{1995TEN, 1998BUO, 2000SOR}}, filtering out the ground state wavefunction (WF) ${\left| {{\Upsilon _0}} \right\rangle }$ from a given trial (or guiding) WF, $\left| {{\Psi _{\text{G}}}} \right\rangle$. Since the eigenstates of the Hamiltonian are a complete basis set, the trial WF can be expanded as:
\begin{equation}
\left| {{\Psi _{\text{G}}}} \right\rangle  = \sum\limits_{i \ge 0} {{a_i}\left| {{\Upsilon _i}} \right\rangle },
\end{equation}
where ${a_i}$ is the coefficient for the $i$-th eigenvectors ${\left|{\Upsilon _i}\right \rangle}$. Therefore, by applying ${\left( {\Lambda  - \hat {\mathcal{H}}} \right)^M}$, one can obtain
\begin{equation}
\begin{split}
\left| {{\Upsilon _0}} \right\rangle &\propto \mathop {\lim }\limits_{M \to \infty } {\left( {\Lambda  - \hat {\mathcal{H}}} \right)^M}\left| {{\Psi _{\text{G}}}} \right\rangle  \\
& = \mathop {\lim }\limits_{M \to \infty } {\left( {\lambda  - {E_0}} \right)^M}\left[ {{a_0}\left| {{\Upsilon _0}} \right\rangle  + \sum\limits_{n \ne 0} {{{\left( {\frac{{\lambda  - {E_i}}}{{\lambda  - {E_0}}}} \right)}^M}{a_i}\left| {{\Upsilon _i}} \right\rangle } } \right],
\end{split}
\label{eq:power-method}
\end{equation}
where ${\boldsymbol \Lambda}$ is a diagonal matrix with ${{\Lambda _{x,x^\prime}}} = \lambda \delta_{x,x^\prime}$ and ${{E_i}}$ is $i$-th eigenvalue of $\hat {\mathcal{H}}$. Assuming a non-degenerate ground state and $\left| \frac{{\lambda  - {E_i}}}{{\lambda  - {E_0}}} \right| < 1 \textrm{~~for~~} i \ge 1$, the projection filters out the ground state WF ${\left|{\Upsilon _0}\right\rangle}$ from a given trial WF $\left| {{\Psi _{\text{G}}}} \right\rangle$, as long as the trial WF is not orthogonal to the true ground state ({\it i.e.}, ${a_0} \equiv \braket{{{\Psi _{\text{G}}}}|{{\Upsilon _0}}} \ne 0$). The condition $\left| \frac{{\lambda  - {E_i}}}{{\lambda  - {E_0}}} \right| < 1$ can be verified on a lattice for a large enough $\lambda$. On the continuum, the spectrum of the \emph{ab initio} $H$ is not bounded from above. Thus, the above condition is verified only in the $\lambda \rightarrow \infty$ limit, which can be easily taken, as shown later on.
This projection scheme is called the power method.
In practice, Eq.~{\ref{eq:power-method}} can be implemented iteratively:
\begin{equation}
\ket{\Psi_{n+1}} = (\Lambda - \mathcal{\hat{H}}) \ket{\Psi_{n}}.
\end{equation}
By expanding over the given basis set with the guiding function $\Psi_{\text{G}}(x)$, we have:
\begin{equation}
\Psi_{\text{G}}(x')\Psi_{n+1}(x') = \sum_x \mathcal{G}_{x',x}\Psi_{\text{G}}(x)\Psi_{n}(x)
\label{eq:master-equation}
\end{equation}
where $\Psi_{n}(x) \equiv \braket{x|\Psi_{n}}$ and
\begin{equation}
\mathcal{G}_{x',x} \equiv \braket{x'|\Lambda - \mathcal{\hat{H}}|x} \cfrac{\braket{x'|\Psi_{\text{G}}}}{\braket{x|\Psi_{\text{G}}}} \equiv \lambda \delta_{x',x} - H_{x',x} \cfrac{\Psi_{\text{G}}(x')}{\Psi_{\text{G}}(x)}
\label{eq:def-greenfunction}
\end{equation}
is the so-called importance sampling Green's function.
Within a statistical implementation of the power method, we would be tempted
to interpret Eq.~\ref{eq:master-equation}
as a master Equation for a stochastic variable, where $\Pi_n(x)\equiv\Psi_G(x)\Psi_n(x)$
%\Psi(x)$
represents the probability distribution at the iteration $n$ and $\mathcal{G}$ is the transition probability. However, some important points must be elucidated in order to reach a final statistical interpretation of this quantum evolution.

%
%\vspace{2mm}
To interpret the Green's function as a transition probability, all matrix elements of the Green's function must be non-negative.
However, it is well known that for Fermions $\mathcal{G}_{x',x}$ is not always positive. This issue is known as the {\it sign problem} and is circumvented via the FNA by defining an effective FN Hamiltonian, as detailed in Appendix~\ref{app:fna_and_lattice_regularization}. In the main text, we assume that all matrix elements are non-negative, but the discussion can be easily generalized to FN cases.

%\vspace{2mm}
Even when the Green's function is non-negative, $\mathcal{G}_{x',x}$ is generally not normalized. Therefore, we need to introduce a normalization factor $b_x = \sum_{x'} \mathcal{G}_{x',x}$ so that we can define the $p$ matrix as
\begin{equation}
p_{x',x} = \frac{\mathcal{G}_{x',x}}{b_x}.
\label{eq:normalization-probablity}
\end{equation}
$p_{x',x}$ can now be interpreted as transition probability since it is non-negative and row-wise normalized ($\sum_{x'} p_{x',x} = 1$). Then, the corresponding Markov process considers the presence of the normalization factor $b_x$ by adding a weight $w$ in the statistical implementation of the projection, which is thus defined by a dyad $(x_n, w_n)$. Its evolution from $(x_n, w_n)$ is described by:
\begin{eqnarray}
x_{n+1} &=& x' \quad \text{with probability } p_{x',x_n}, \label{eq:projection-1} \\
w_{n+1} &=& w_n \cdot b_x. \label{eq:projection-2}
\end{eqnarray}
The Markov process, comprising the diffusion of the configuration as well as its weight, can be cast in the following master Equation:
\begin{equation}
\mathcal{P}_{n+1}(x', w') = \sum_x \int dw \, K(x', w' | x, w) \, \mathcal{P}_n(x, w),
\label{eq:kernel_equation}
\end{equation}
where the transition kernel $K$ is defined as
%
%\begin{equation}
$K(x', w' | x, w) = p_{x',x} \, \delta(w'-w b_x)$.
%\end{equation}
%
Multiplying both sides of Eq.~\ref{eq:kernel_equation} by $w'$ and integrating over it yields:
%
%\begin{equation}
%\int dw' \, w' \mathcal{P}_{n+1}(x', w') = \sum_x \mathcal{G}_{x',x} \int dw \, w \mathcal{P}_n(x, w)
%= \sum_x p_{x',x} b_x \int dw \, w \mathcal{P}_n(x, w).
%\end{equation}
\begin{equation}
\int dw' \, w' \mathcal{P}_{n+1}(x', w') 
= \sum_x p_{x',x} b_x \int dw \, w \mathcal{P}_n(x, w)
= \sum_x \mathcal{G}_{x',x} \int dw \, w \mathcal{P}_n(x, w).
\end{equation}
The marginal weighted probability is nothing but the mixed many-body distribution, i.e., $\int dw \, w \, \mathcal{P}_n(x,w) = \Pi_n(x)$ at step $n$.
Thus, the desired ground-state energy $E_0$ is computed at the projection step $n$ as
\begin{equation}
E_0 = \frac{\langle \Psi_G | \hat{H} | \Psi_0 \rangle}{\langle \Psi_G | \Psi_0 \rangle}
= \frac{\sum_x e_L(x) \Psi_G(x) \Psi_n(x)}{\sum_{x'} \Psi_G(x') \Psi_n(x')},
\end{equation}
where the local energy, $e_L(x)$, is defined as
\begin{equation}
e_L(x) = \frac{\langle \Psi_G | \hat{H} | x \rangle}{\langle \Psi_G | x \rangle}
= \sum_{x'} H_{x,x'} \cdot \frac{\Psi_G(x')}{\Psi_G(x)}.
\end{equation}
%

%\vspace{2mm}
To implement the Markov process in a computable scheme, we define its basic element, the so-called \emph{walker}, determined by the dyad ($x$, $w$). Note how the weight $w$ is associated to the amplitude of the mixed distribution at $x$. The walker changes its configuration and weight by performing a Markovian process with a discrete iteration time $n$: the dyad ($x_n$, $w_n$) is distributed according
to $\mathcal{P}_n(x, w)$.
Most importantly, the walker determines, in a statistical sense, the mixed quantum state $\Psi_G(x) \Psi_n(x)$ as:
\begin{equation}
\Psi_G(x) \Psi_n(x)
= \int dw \, w \, \mathcal{P}_n(x,w)
\sim \braket{\delta_{x,x_n}
%e_L(x_n) 
w_n},
\label{eq:mixed-quantum-distribution}
\end{equation}
where $\braket{\cdots}$ denotes the statistical average over {\emph{independent}} Markov chains at the $n$-th step.
Thus, the desired ground-state energy $E_0$ is computed as
\begin{equation}
E_0 = \frac{\sum_x e_L(x) \Psi_G(x) \Psi_n(x)}{\sum_{x'} \Psi_G(x') \Psi_n(x')} = \frac{\sum_x e_L(x) \int dw \, w \, \mathcal{P}_n(x,w)}{\sum_{x'} \int dw' \, w' \, \mathcal{P}_n(x',w')} \sim \cfrac{\braket{e_L(x_n) w_n}}{\braket{w_n}}.
\label{eq:ground_state_energy_standard}
\end{equation}
However, in practice, the strategy of simply computing independent Markov chains and averaging their results is not adopted. This is because walkers with extremely large or small weights may appear, leading to numerical instability. While keeping a multi-walker strategy with $N_w$ walkers, we instead reconfigure their weights via the so-called branching technique for a more efficient computation~{\cite{2017BEC}}. Every $N_{\rm proj}$ projections, the code performs a branching step as follows:
\begin{enumerate}
\item[(1)] Select the new walkers from the previous ones with a probability that is proportional to the former walkers' weights before the branching step:
\begin{equation}
{p_{\alpha ,n}} = \frac{{{w_{\alpha ,n}}}}{{\sum\limits_\beta  {{w_{\beta ,n}}} }},
\label{eq:branching-2}
\end{equation}
%
%Note how in a multi-walker formalism the dyad acquires the additional walker index $\alpha$: $(x_{\alpha, n}, w_{\alpha, n})$.
%
\item[(2)] Set the new weights equal to the weights average of the former walkers:
\begin{equation}
w_{\alpha, n+1} = \bar w_n \equiv \frac{1}{{{N_w}}}\sum\limits_\beta  {{w_{\beta ,n}}} ~~\forall \alpha.
\label{eq:branching-1}
\end{equation}
\end{enumerate}
Note that, in the multi-walker formalism, the dyad acquires the additional walker index $\alpha$: $(x_{\alpha, n}, w_{\alpha, n})$.
As rigorously proved in Refs.~{\onlinecite{1998BUO, 2017BEC}}, this walkers reconfiguration does not change their statistical average, and suppresses their fluctuations by dropping (spawning) walkers having small (large) weights.
The selection of new configurations in a branching step is implemented by an efficient reconfiguration scheme deciding which walker survives or dies by extracting $N_w$ {\emph{correlated}} random numbers~\cite{2017BEC}:
\begin{equation}
Z_{\alpha} \;=\;\frac{\xi + \alpha - 1}{N_w},\quad \alpha=1,\dots,N_w,
\label{eq:correlated_random_numbers}
\end{equation}
where $\xi$ is a uniform variate 
in $[0,1)$.  This set of numbers is then used to select the new configuration by comparing it with the normalized weights $w_{\alpha ,n}/\bar w_n$. This reconfiguration provides the same stabilization effect as the conventional branching scheme~\cite{2001FOU},
%~\cite{2001FOU, 2017BEC}, 
but with the advantage that the total number of walkers remains fixed throughout the simulation.
%~{\cite{2017BEC}}.

%
%\vspace{2mm}
To make full use of the available samples, observables are evaluated using all samples obtained after the Markov chain has reached equilibrium. After a thermalization time $p$ expressed in branching step units, the configurations $x$ along the Markov process will be equilibrated according to $\pi_{\rm eq}(x)$; then, at each branching step $n$ we can imagine to start the projection $n-p$ steps backwards with initial weights $w_{\alpha,n-p} = 1 ~~\forall \alpha$. In this case, the initial probability is given by $\mathcal{P}_{n-p}(x, w) = \delta(w - 1) \pi_{\rm eq}(x)$. To perform the Monte Carlo averages with the branching technique, we can accumulate the average weights in the global population factor ${\bar{G}}_n^p$ developed after $n-p$ branching steps:
\begin{equation}
{\bar{G}}_n^p = \prod\limits_{j = 1}^p {{{\bar w}_{n - j}}},
\label{eq:many-walker-weight-conventional-lrdmc}
\end{equation}
while setting the weights to 1 after each walkers reconfiguration. In this way, the ground state energy is computed as:
\begin{equation}
{E_0} \approx \frac{{\sum\nolimits_n {{\bar{G}}_n^p{\bar{e}_L}\left( {{x_n}} \right)} }}{{\sum\nolimits_n {{\bar{G}}_n^p} }},
\label{eq:ground-state-energy-estimation-weighted}
\end{equation}
where ${\bar{e}_L}\left( {{x_n}} \right)$ is the mean local energy averaged over the walkers, which reads:
\begin{equation}
{\bar{e}_L}\left( {{x_n}} \right) = \frac{{\sum\nolimits_\alpha  {{w_{\alpha ,n}}{e_L}\left( {{x_{\alpha ,n}}} \right)} }}{{\sum\nolimits_\alpha  {{w_{\alpha ,n}}} }},
\label{eq:local-energy-weighted}
\end{equation}
and is evaluated just before each reconfiguration, with weights $w_{\alpha ,n}$ acquired only during the evolution after the $n-1$-th branching step.

% 
%\vspace{2mm}
When applying the GFMC in the context of \emph{ab initio} calculation, the original continuous Hamiltonian is regularized by allowing electron hopping with step size $a$. The corresponding Hamiltonian ${{\hat{\mathcal{H}}}^a}$  is then defined such that ${{{\hat {\mathcal{H}}}^a}} \to {\hat {\mathcal{H}}}$ for $a \to 0$. 
After discretizing the space through a lattice spacing $a$, the hopping elements $H_{x',x}$ of the Hamiltonian can be computed by considering only the hoppings of \emph{each electron} to its neighboring grid points in the current electronic configuration. As a result, $H_{x',x}$ becomes sparse, allowing the GFMC method to be applied to a continuous Hamiltonian. This approach is referred to as LRDMC~{\cite{2005CAS}}. Then, the energy obtained by LRDMC exhibits a bias of the order of $\mathcal{O}(a^2)$ with respect to $a$, and therefore, to access unbiased total energies (or other observables), LRDMC calculations are performed for several values of $a$, and an extrapolation to $a \to 0$ is carried out.
The technical details are explained in the Appendix~\ref{app:fna_and_lattice_regularization}.

\subsection{The conventional LRDMC algorithm and its intrinsic load-imbalance}
\label{subsec:load-imbalance-conventional-lrdmc}
As we have seen above, in the LRDMC framework, the projection of the wavefunction is governed by the repeated application of a Green's function, interpreted as a stochastic transition matrix (Eqs.~{\ref{eq:normalization-probablity},
{\ref{eq:projection-1}}, {\ref{eq:projection-2}}}). To ensure a probabilistic interpretation, the Green's function must be non-negative, which often requires shifting the Hamiltonian by a constant $\Lambda$ such that all diagonal elements become positive (Eq.~{\ref{eq:def-greenfunction}}). While on a lattice this shift can always be applied, on the continuum this is prevented from the fact that the local Hamiltonian terms can become arbitrarily large, due to the divergence of the Coulomb potential. Moreover, choosing an excessively large shift degrades the efficiency, as it increases the probability for a walker to remain in the same configuration. This follows from $\mathcal{G}_{x,x} \gg \mathcal{G}_{x' \ne x,x}$, thus $p_{x,x} \gg p_{x' \ne x,x}$, leading to long autocorrelation times and inefficient sampling.

%\vspace{1mm}
To circumvent this issue, the projection steps (Eqs.~{\ref{eq:projection-1}} and {\ref{eq:projection-2}}) are reformulated in the $\lambda \rightarrow \infty$ limit, so defining a {\it continuous-time} stochastic process. Rather than evolving the walker with discrete steps governed by the bare $p_{x',x}$ (Eq.~{\ref{eq:normalization-probablity}}), the algorithm considers the number of successive diagonal moves {\emph{stochastically}} followed by a single off-diagonal move driven by $p_{x' \ne x,x}$.
Within this formalism, the number $k$ of successive diagonal moves is drawn from a uniformly distributed random number $\xi \in [0,1)$, resulting in~{\cite{trivedi1990ground,2000SOR}}:
%{{\cite{2005CAS, 2017BEC}}:
%
\begin{equation}
%k = \left\lfloor \frac{\ln(1 - \xi)}{\ln(1 - p_{x,x})} \right\rfloor.
k = \left\lfloor - \frac{\ln(1 - \xi)}{1-p_{x_n,x_n}} \right\rfloor.
\label{eq:k-times-there}
\end{equation}
The corresponding walker weight is updated as:
\begin{equation}
w_{n+1} = w_n \cdot b^k.
\label{eq:update-walker-weight-discretized}
\end{equation}
This approach is generalized to the continuous-time limit, namely, $\lambda \rightarrow \infty$, i.e. with an infinitesimal time step defined as $\delta \tau \equiv 1/\lambda \rightarrow 0$~{\cite{casula2005PhD,2017BEC}}. In this limit, the projection in Eq.~{\ref{eq:power-method}}, ${\left( {\Lambda  - \hat {\mathcal{H}}} \right)^M}$, is evaluated by keeping $M/\lambda=\tau$ finite, such that the imaginary time evolution is cast in $\exp \left( { - \tau \hat {\mathcal{H}}} \right)$, apart from an irrelevant constant $\lambda^M$.
At the same time, the denominator of Eq.~\ref{eq:k-times-there} can be written as:
\begin{equation}
1-p_{x,x} = \sum_{x' (\ne x)} p_{x',x} = \frac{\sum_{x' (\ne x)} \mathcal{G}_{x',x}}{\lambda - e_L(x)} \rightarrow \delta \tau \sum_{x' (\ne x)} \mathcal{G}_{x',x} + O\left((\delta \tau)^2\right).
\label{eq:small_tau_limit}
\end{equation}
By plugging Eq.~\ref{eq:small_tau_limit} into Eq.~\ref{eq:k-times-there}, one obtains at the leading order in $\delta \tau$:
\begin{equation}
{\tau _\xi} = - \frac{\ln \left( 1 - {\xi} \right)}{\sum_{x' (\ne x_n)} \mathcal{G}_{x',x_n}},
\label{eq:tau-xi}
\end{equation}
which corresponds to $\tau _\xi = k ~ \delta\tau$, the persistence time at configuration $x$ in the continuous-time limit. Correspondingly, Eq.~\ref{eq:update-walker-weight-discretized} becomes:
\begin{equation}
w_{n+1} = w_n \cdot \exp \left( { - {\tau _\xi }{e_L}\left( x_n \right)} \right).
\label{eq:weight-update-conventional-lrdmc}
\end{equation}
Therefore, in the continuous-time limit formulation, one estimates the persistence time at configuration $x_n$ via Eq.~\ref{eq:tau-xi} and updates the walkers weights according to Eq.~\ref{eq:weight-update-conventional-lrdmc}, before performing an off-diagonal move \emph{exclusively} driven by the off-diagonal elements $\mathcal{G}_{x',x_n}$ appropriately normalized with $\sum_{x' (\ne x_n)} \mathcal{G}_{x',x_n}$.
%
%The update of the walker weight (Eq.~{\ref{eq:update-walker-weight-discretized}}) is replaced with:
%
%\begin{equation}
%w \to w\exp \left( { - {\tau _\xi }{e_L}\left( x \right)} \right),
%\label{eq:weight-update-conventional-lrdmc}
%\end{equation}
%
%and 
The imaginary time is then updated according to ${\tau _{{\text{left}}}} \leftarrow {\tau _{{\text{left}}}} - {\tau _\xi }$ at each off-diagonal update until ${\tau _{{\text{left}}}}$ becomes 0. 
%The update of $\tau$ at each step is a diagonal move time step determined by a uniform random number $0 \le \xi < 1$ as:
%
%\begin{equation}
%{\tau _\xi} = - \log \left( 1 - {\xi} \right)/{b_x}.
%\label{eq:tau-xi}
%\end{equation}
%
Each branching (reconfiguration) step is performed after an integrated imaginary time evolution $\tau$ that is an input parameter defined \emph{a priori}.
Importantly, this continuous-time formalism allows simulations of the imaginary-time propagator $\exp(-\tau H)$ without requiring the Suzuki-Trotter decomposition, thereby eliminating the systematic Suzuki-Trotter error inherent in traditional DMC implementations. The price to pay is the lattice discretization bias, that has to be extrapolated, as we have seen in the previous Subsection.~{\cite{2005CAS, 2017BEC}}.
The pseudocode of the LRDMC algorithm is shown in Table~\ref{alg:LRDMC-tau}. 

%\subsection{The intrinsic load-imbalance in the conventional LRDMC implementation}

%\vspace{2mm}
The conventional LRDMC algorithm faces a specific challenge in massively parallel environments, particularly with many walkers. In the conventional algorithm, the projection time evolution is governed by stochastic sampling the propagation intervals, or persistence times $\tau_{\xi}$, using uniform variates (Eq.~{\ref{eq:tau-xi}}). As a result, the computational effort required per walker to complete the total time evolution $\tau$ between two branching steps becomes unbalanced: the number of steps necessary for each walker to reach $\tau_{\rm left} = 0$ varies randomly. This imbalance poses a problem because the branching process must wait until {\emph{all}} walkers have completed their projection steps. Consequently, walkers that finish earlier have to remain idle until the slowest walker completes its projection. Consequently, the parallel efficiency of the conventional LRDMC algorithm deteriorates as the number of walkers increases. 
%
%{\vspace{2mm}}
This degradation can be shown to grow logarithmically with the number of walkers $N_w$, , as detailed in Appendix~{\ref{app:emulation}}. Indeed, the maximum number of moves during a fixed time interval between two branching steps increases as $\log(N_w)$, indicating that the actual computation time also increases as $\log(N_w)$, even though the per-walker workload remains unchanged on average.

%{\vspace{2mm}}
One might consider to alleviate the load-imbalance problem within the framework of the conventional LRDMC algorithm, by choosing a sufficiently large projection time $\tau$ between two consecutive branching steps. However, employing a large value of $\tau$ has two main limitations: a degradation of computational efficiency and a possible numerical overflow arising from the accumulated walker weights.
Regarding the first issue, if $\tau$ is taken too large, the weight distribution among walkers becomes highly imbalanced. As a result, at the branching step, a substantial number of walkers are eliminated, and the walkers' survival rate diminishes. This, in turn, decreases the effective sample size, thereby increasing the statistical error. Consequently, a too large $\tau$ severely undermines the efficiency of the LRDMC calculation, thus a tradeoff must be found between the load-imbalance reduction and the increase of the statistical error bar.
As for the second issue, in the conventional LRDMC algorithm, the weights are updated according to Eq.~\ref{eq:weight-update-conventional-lrdmc}, and as evident from this expression, they grow exponentially at each step. Therefore, when a large $\tau$ is used, some weights could lead to numerical overflow. Furthermore, the accumulation of weights using Eq.\ref{eq:many-walker-weight-conventional-lrdmc} can become infeasible for the same reason.

\vspace{2mm}
To overcome the load-imbalance of the conventional LRDMC algorithm, we propose a new LRDMC algorithm that ensures a perfectly load-balanced evolution of walkers between two branching steps, independently of their number. The details of the new algorithm are described in the next Section.

\subsection{The load-balanced LRDMC algorithm}
\label{subsec:load-balanced-lrdmd}
%\vspace{1mm}
The idea behind a load-balanced LRDMC algorithm relies on the possibility to define a branching rate determined by a fixed number of off-diagonal moves, set equal for every walker, rather than by the persistence time, which depends instead on each walker path, finally responsible of the load imbalance issue. In Sec.~\ref{subsec:load-imbalance-conventional-lrdmc}, we have seen that the persistence time is proportional to the magnitude of the diagonal Green's function elements of the LRDMC Hamiltonian. Thus, we need to define a projection process that is driven by an \emph{auxiliary} Hamiltonian with zero diagonal matrix elements, in such a way that the corresponding walkers will never stop, i.e. they will constantly hop from site to site. This can be realized with the strategy described as follows. Given the FN LRDMC Hamiltonian $H$ on a lattice, we can formally divide its matrix elements in off-diagonal $\bar{\mathcal{H}}_{x',x}$ and diagonal $\mathcal{W}(x) \equiv \mathcal{H}_{x,x}$ ones, such that:
\begin{equation}
\mathcal{H}_{x',x} = \bar{\mathcal{H}}_{x',x} + \delta_{x,x^\prime} \mathcal{W}(x),
\label{eq:H_eff_0local}
\end{equation}
where $\bar{\mathcal{H}}_{x,x}=0$ by construction. Notice that the above partition can be realized $\forall x$ thanks to the potential regularization provided by the definition of the LRDMC Hamiltonian, which cuts off any divergence of the bare \emph{ab initio} potential by means of the length scale $a$ (see Appendix \ref{app:fna_and_lattice_regularization} for a detailed description of the regularized potential). From Eq.~\ref{eq:H_eff_0local}, it follows that the wavefunction $\Psi_0$, ground state of the FN LRDMC Hamiltonian with energy $E_0$, will fulfill the following Schr\"odinger equation:
\begin{equation}
 \sum_{x}\bar{\mathcal{H}}_{x',x} \Psi_0(x)=\left(E_0 - \mathcal{W}(x') \right) \Psi_0(x').
\end{equation} 
In analogy with the original Schr\"odinger equation, one can define a modified Green function (with importance sampling) based on $\bar{\mathcal{H}}_{x',x}$:
\begin{equation}
\mathcal{G}_{x^\prime,x} = -{1 \over \mathcal{W}(x) - E_0} \bar{\mathcal{H}}_{x',x} \cdot \cfrac{\Psi_G(x')}{\Psi_G(x)}.
\label{eq:modified_greens_function}
\end{equation}
According to Eq.~\ref{eq:modified_greens_function}, one can implement a stochastic process, in close analogy with what has been derived for conventional LRDMC algorithm based on Eq.~\ref{eq:def-greenfunction}
by using the decomposition:
\begin{eqnarray}
 \mathcal{G}_{x^\prime,x} &=& b_x p_{x^\prime,x}, \nonumber \\
 b_x &=& {1 \over \mathcal{W}(x)-E_0}  \bar b_x,  \nonumber \\
 \bar b_x &=&-\sum\limits_{x^\prime } \Psi_G(x^\prime) \bar{H}_{x^\prime,x}/\Psi_G(x),  \nonumber \\
 p_{x^\prime,x} &=&-{\Psi_G(x^\prime) \bar{H}_{x^\prime,x}/\Psi_G(x) \over \bar b_x}.
\label{eq:greens_function_partition_load_balance}
\end{eqnarray}
In order to interpret $b_x$ as a legitimate weight while $p_{x^\prime,x}$ is the normalized transition probability, $b_x$ must be non negative in the FNA. This follows from the positiveness of $\mathcal{W}(x)-E_0 ~~\forall x$. To show this, take the minimum of the potential 
$\mathcal{W}(x_0)$ occurring for some configuration $x_0$. Since the trial state 
$\psi_T(x)=\delta_{x,x_0}$ has variational energy $\mathcal{W}(x_0)$, this implies immediately that $V(x)-E_0>0 ~~\forall x$.
Then, as we have already seen in Subsec.~\ref{subsec:gfmc-and-lrdmc}, the algorithm built on the set of Eqs.~\ref{eq:greens_function_partition_load_balance}, performs the conventional lattice update:
\begin{eqnarray}
x_{n+1} &=& x' {\rm ~~ with~~} p_{x',x}, \nonumber \\
w_{n+1} &=& w_n b_x.
\label{eq:lattice_update_load_balance}
\end{eqnarray}
A master equation formally equivalent to Eq.~\ref{eq:master-equation} can then be drawn for the above stochastic process. One can show (see Appendix \ref{app:stationary_distribution_load_balance}) that a \emph{stationary} distribution of this process is the marginal weighted probability given by:
\begin{equation}
\int dw w P_n(x,w) = \left( \mathcal{W}(x)-E_0 \right) \Psi_G(x)\Psi_0(x),
\label{eq:stationary_load_balance}
\end{equation}
which is the conventional LRDMC mixed distribution (Eq.~\ref{eq:mixed-quantum-distribution}) modulated by the factor $\left( \mathcal{W}(x)-E_0 \right)$.
%
%where $e_L(x)=-\bar b_x + \mathcal{W}(x)$ is the local energy.
%
As derived in the Appendix \ref{app:stationary_distribution_load_balance}, we can prove that for a good enough mixed distribution $\left( \mathcal{W}(x)-E_0 \right) \Psi_G(x)\Psi(x)$, namely with $\Psi(x)$ close enough to the true FN ground-state wavefunction $\Psi_0(x)$, the initial state converges to the final distribution in Eq.~\ref{eq:stationary_load_balance} for $M \to \infty$ with the modified Green function in Eq.~\ref{eq:modified_greens_function}.
%
%\begin{equation} \label{prob}
%\Pi(x) = \left( \mathcal{W}(x) - E_0 \right)  \Psi_G(x) \Psi_0(x).
%\end{equation}
%
Owing to these convergence properties, for averaging an observable we need to consider the additional factor $\left( \mathcal{W}(x)-E_0 \right)$ in the distribution, which must be compensated by an extra inverse factor in the weights. 
%With the modification of the Green function and the importance sampling:
%
%\begin{equation}
% \left( \mathcal{W}(x)-E_0 \right) \Psi_G(x)\Psi_n(x) \equiv \int dw w P_n(x,w) = \braket{w_n \delta_{x,x_n}},
%\end{equation}
%
For instance, the total energy is evaluated as:
\begin{equation}
E_0 = \cfrac{\sum_x e_L(x) \Psi_T(x)\Psi_n(x)}{\sum_x'\Psi_T(x')\Psi_n(x')},
\end{equation}
which can be computed by
\begin{equation}
E_0 \sim \cfrac{\left\langle e_L(x_n) \cfrac{w_n}{\mathcal{W}(x_n)-E_0}\right\rangle}{\left\langle\cfrac{w_n}{\mathcal{W}(x_n)-E_0}\right\rangle},
\label{eq:ground_state_energy_load_balance}
\end{equation}
where $e_L(x)=-\bar b_x + \mathcal{W}(x)$ is the local energy of the FN LRDMC Hamiltonian $\mathcal{H}_{x',x}$ in Eq.~\ref{eq:H_eff_0local}, and $\braket{\cdots}$ is the statistical average over independent Markov chains at the $n$-th step, denoted as $(x_n, w_n)$, and distributed according to $P_n(x,w)$ (Eq.~\ref{eq:stationary_load_balance}).
%(c.f., it is replaced with samplings over MCMC steps in practice, as explained in Sec.~{\ref{subsec:gfmc-and-lrdmc}}), 
%and $E_0$ is the ground state energy. 
From Eq.~\ref{eq:ground_state_energy_load_balance}, it is clear that the ground state estimate
$E_0$, which is the value appearing in the modified Green's function of Eq.~\ref{eq:modified_greens_function}, can be self-consistently determined by averaging the local energy. 

\vspace{2mm}
The sampling strategy implied by $\braket{\cdots}$ in Eq.~\ref{eq:ground_state_energy_load_balance} and adopted by the load-balanced LRDMC algorithm, follows exactly the same multi-walkers framework described in Subsec.~\ref{subsec:gfmc-and-lrdmc}, together with the same walkers reconfiguration scheme (see Eqs.~\ref{eq:branching-2}--\ref{eq:ground-state-energy-estimation-weighted}, which perfectly hold also in this case). A difference between the conventional and the load-balanced LRDMC algorithms shows up in the weighted averages of the multi-walker framework. Indeed, the mean local energy averaged over the walkers, corresponding to Eq.~\ref{eq:local-energy-weighted} in the conventional algorithm, becomes now 
\begin{equation}
{\bar{e}_L}\left( {{x_n}} \right) = \cfrac{{\sum\nolimits_\alpha  {{\cfrac{{w_{\alpha ,n}}}{\mathcal{W}(x_{\alpha,n}) - E_0}} \cdot {e_L}\left( {{x_{\alpha ,n}}} \right)} }}{{\sum\nolimits_\alpha {\cfrac{{w_{\alpha ,n}}}{\mathcal{W}(x_{\alpha,n}) - E_0}} }},
\label{eq:local-energy-weighted-load-balance}
\end{equation}
where the weights accumulated according to Eqs.~\ref{eq:greens_function_partition_load_balance} and \ref{eq:lattice_update_load_balance} must acquire an extra factor $1/(\mathcal{W}(x) - E_0)$ in the averages to compensate for the modified distribution of Eq.~\ref{eq:stationary_load_balance}. This is the same difference that links Eq.~\ref{eq:ground_state_energy_load_balance} with Eq.~\ref{eq:ground_state_energy_standard}.
In the load-balanced LRDMC algorithm, $E_0$ is updated after each branching step.

\vspace{2mm}
Besides the differences in the weighted averages, the main change with respect to the conventional LRDMC algorithm concerns the way the branching rate is set. Indeed, in this modified LRDMC algorithm, the number of steps between two branchings, denoted $N_{\rm proj}$, is no longer stochastically determined as in the conventional algorithm. In the conventional LRDMC algorithm, as described in Subsec.~\ref{subsec:load-imbalance-conventional-lrdmc}, $N_{\rm proj}$ corresponds to the randomly determined number of projections required to reach the projection time $\tau$ between two branching steps. 
Instead, in the load-balanced LRDMC $N_{\rm proj}$ is defined \emph{a priori} as a fixed input parameter. Indeed, no continuous-time limit is needed, as no $\lambda \rightarrow \infty$ limit is taken. The electrons instantaneously jump from site to site, with a sequence determined by $N_{\rm proj}$, equally set for all walkers.
As a result, the computational cost associated with the projection steps becomes strictly uniform across all walkers. This uniformity is expected to prevent any degradation in parallel efficiency, even when a large number of walkers is employed.
Indeed, as we will demonstrate later, the weak-scaling results clearly show that the conventional algorithm suffers from a decline in weak-scaling performance which goes like $\ln(N_w)$. In contrast, the new algorithm exhibits excellent weak-scaling behavior. This indicates that the proposed modification is crucial for maintaining a high computational throughput, especially in LRDMC calculations that aim to leverage GPUs hardware architectures.
Moreover, we have numerically verified to high precision that both the conventional and load-balanced algorithms yield consistent results in the limit $a \to 0$, provided the average number of projection steps is the same. This confirms the equivalence of the two algorithms in our testbed systems, as far as the convergence properties are concerned.

\vspace{2mm}
The pseudocode of the load-balanced LRDMC algorithm is reported in Table~{\ref{alg:LRDMC-bra}}.

%%%%%%%%%%%%%%%%%%%%%%%%%%%%%%%%%%%%%%%%%%%%%%%%%
% Validation
%%%%%%%%%%%%%%%%%%%%%%%%%%%%%%%%%%%%%%%%%%%%%%%%%
\section{Validation of the algorithm and its implementation}
\label{validation}

We first demonstrate that the conventional algorithm of LRDMC and the load-balanced vesion yield identical results.
In the QMC community, recent efforts have aimed at systematically evaluating the reproducibility of results across independently developed codes~{\cite{2025PIA}}. A notable example is a collaborative study involving 11 different community-developed QMC packages, in which the total energy and binding energy of a water–methane dimer were computed using a standardized basis set and a common effective core potential, thereby allowing for a controlled comparison of code implementations.
In the present work, we adopt the same water–methane system as our benchmark case. The two LRDMC algorithms—conventional and load-balanced—were implemented using the jQMC~{\cite{jQMC}} and TurboRVB~{\cite{2020NAK} packages.
For both oxygen and hydrogen atoms, we employed correlation-consistent effective core potentials (ccECPs)~{\cite{2017BEN, 2018BEN}} and used the associated \texttt{cc-pVTZ} basis sets, as in the aforementioned benchmark study. The angular part of the basis functions was expressed using spherical (solid) harmonics. The trial wavefunction was generated using PySCF~{\cite{2018SUN, 2020SUN}} and converted into the jQMC and TurboRVB wavefunction formats via TREX-IO~{\cite{2023POS}}. 
We adopted the same functional form and variational parameters of the Jastrow factor used specifically in the TurboRVB calculations of the benchmark study~\cite{2025PIA} for both the jQMC and TurboRVB calculations.
A central technical challenge in DMC calculations using ECPs is the treatment of the non-local part of the ECP operator. In this work, we adopt the Determinant Locality Approximation (DLA)~{\cite{2019ZEN}} within the LRDMC framework, which corresponds to using the DLA with the T-move~{\cite{2006CAS}} scheme in standard DMC~{\cite{2024NAK}}.

\vspace{2mm}
To begin with, we verified that the conventional and load-balanced LRDMC algorithms yield identical results for the water molecule, which is as one of the benchmark systems in the previous reproducibility study~{\cite{2025PIA}}. Table~{\ref{tab:comparison_water_lrdmc_energies}} shows the total energies computed with both algorithms for $a$ = 0.05, 0.10, 0.15, 0.20, 0.25, and 0.30 a.u. This verification was carried out using both jQMC and TurboRVB packages. The conventional LRDMC results obtained with TurboRVB were taken from Ref.~{\onlinecite{2025PIA}}, while all other data were newly computed in the present work. As shown in Table~{\ref{tab:comparison_water_lrdmc_energies}}, for all values of $a$, the total energies obtained with both jQMC and TurboRVB agree within the error bars between the conventional and load-balanced LRDMC algorithms. This demonstrates that, given the same trial wavefunction, both algorithms sample the same fixed-node ground state at fixed $a$, confirming the numerical equivalence of the two projection algorithms. A more detailed discussion of the computational performance and parallelization efficiency (i.e. weak-scaling) of these algorithms is provided in the next Section.

\vspace{2mm}
Next, we compare the DMC and LRDMC results obtained from different packages. As demonstrated in Ref.~{\onlinecite{2025PIA}}, the energies at finite time-step discretization ($\tau$) in DMC and finite lattice-space discretization ($a$) in LRDMC are dependent on the details of each package's implementation, and thus are not suitable for direct comparison. Therefore, we perform calculations analogous to those in the reproducibility study~{\cite{2025PIA}} to assess the consistency of our implementation. Specifically, for the methane molecule, water molecule, and methane-water dimer, we compute the total energies in the $a \rightarrow 0$ limit using the load-balanced LRDMC implementation in both jQMC and TurboRVB. These results are then compared with the $\tau \rightarrow 0$ DMC energies reported in Ref.~{\onlinecite{2025PIA}}, obtained using QMCPACK~{\cite{2018KIM, 2020KEN}} and CASINO~{\cite{2020NEE}}, as well as with the $a \rightarrow 0$ extrapolated values from the conventional LRDMC implementation in TurboRVB~{\cite{2020NAK}}.
The results are summarized in Fig.~{\ref{fig:inter-software-comparison}} and Table~{\ref{tab:inter-software-comparison}}.
As clearly shown in these comparisons, the computed total energies and binding energies exhibit excellent agreement with the literature values, thereby validating the correctness of our load-balanced LRDMC implementations in both jQMC and TurboRVB packages.

%%%%%%%%%%%%%%%%%%%%%%%%%%%%%%%%%%%%%%
% Figure
\begin{figure*}
    \centering
    \includegraphics[width=1.0\linewidth]{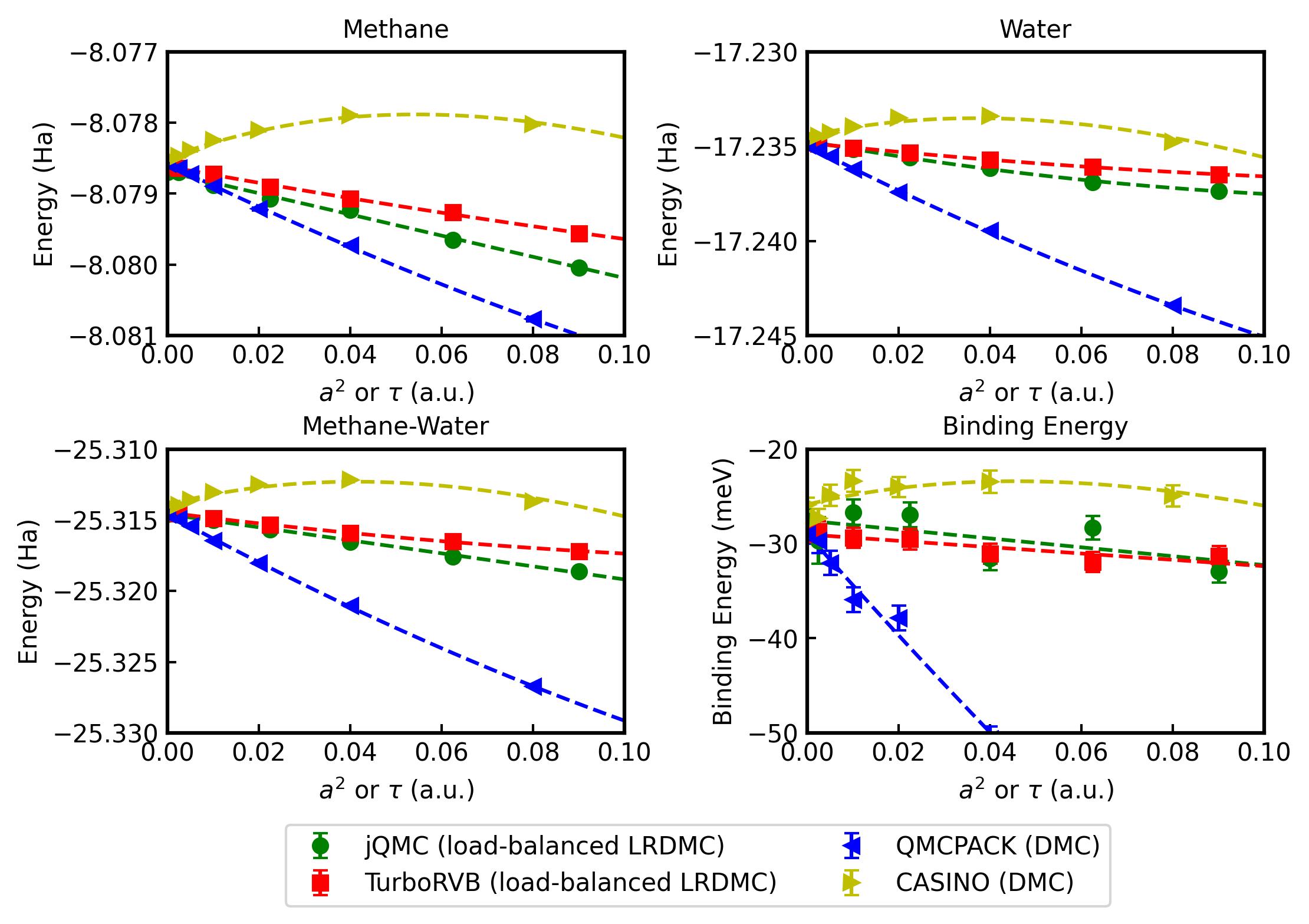}
    \caption[]{The total energy of Methane, Water, and Methane-Water dimer, and the binding energy. The data for QMCPACK, and CASINO are taken from Ref.~{\onlinecite{2025PIA}}. The error bars in the graphs represent one standard deviation (1$\sigma$). The corresponding numbers are shown in Table~{\ref{tab:inter-software-comparison}}.
    }
    \label{fig:inter-software-comparison}
\end{figure*}
%%%%%%%%%%%%%%%%%%%%%%%%%%%%%%%%%%%%%%

%%%%%%%%%%%%%%%%%%%%%%%%%%%%%%%%%%%%%%
\begin{table}
\caption{LRDMC energies of the water molecule for discretized lattice spaces ($a$), computed using both the load-balanced and conventional algorithms implemented in jQMC and TurboRVB.}
\label{tab:comparison_water_lrdmc_energies}
\begin{tabular}{c|cc|cc}
\Hline
\multirow{3}{*}{$a$ (bohr)} 
  & \multicolumn{4}{c}{LRDMC Energy (Ha)}             \\ 
\cline{2-5}
  & \multicolumn{2}{c|}{jQMC} 
    & \multicolumn{2}{c}{TurboRVB}               \\ 
\cline{2-5}
  & load-balanced   & conventional   
    & load-balanced   & conventional       \\ 
\Hline
0.05 & -17.23483(3) & -17.23493(5) & -17.23485(2) & -17.23487(1) \\
0.10 & -17.23516(2) & -17.23514(4) & -17.23509(2) & -17.23507(1) \\
0.15 & -17.23559(3) & -17.23562(5) & -17.23535(2) & -17.23540(1) \\
0.20 & -17.23616(2) & -17.23622(5) & -17.23571(2) & -17.23573(1) \\
0.25 & -17.23688(2) & -17.23680(6) & -17.23608(2) & -17.23611(1) \\
0.30 & -17.23735(2) & -17.23723(5) & -17.23648(2) & -17.23648(1) \\
\Hline
\end{tabular}
\end{table}

%%%%%%%%%%%%%%%%%%%%%%%%%%%%%%%%%%%%%%

%%%%%%%%%%%%%%%%%%%%%%%%%%%%%%%%%%%%%%
\begin{table*}[t]
  \centering
  \caption{\label{tab:inter-software-comparison} The total energies (Ha) of the Methane, Water, and Methane-Water, and its binding energy (meV), obtained with the extrapolation to $\tau \rightarrow 0$ (DMC) and $a \rightarrow 0$ (LRDMC). The CASINO (DMC), QMCPACK (DMC), and TurboRVB (Conventional LRDMC) results are taken from Ref.{\onlinecite{2025PIA}}. Numbers in parentheses indicate 1$\sigma$ in the last digit(s).}
  \begin{tabular}{l|ccc|c}
    \Hline
    Package               & Methane (Ha)  & Water (Ha)    & Methane–Water (Ha) & Binding energy (meV) \\
    \Hline
    CASINO (DMC)        & -8.07856(1)   & -17.23473(2)  & -25.31432(3)       & -26.8(1.0)           \\
    QMCPACK (DMC)           & -8.07858(2)   & -17.23482(3)  & -25.31443(7)       & -29.0(1.1)           \\
    %CMQMC (DMC) & -8.07848(2) & -17.23460(1) & -25.31411(3) & -27.8(3) \\ 
    TurboRVB (Conventional LRDMC)  & -8.07860(1)   & -17.23479(1)  & -25.31445(1)       & -28.1(3)           \\
    \hline
    TurboRVB (Load-balanced LRDMC)  & -8.07862(2)   & -17.23482(2)  & -25.31447(2)       & -29.0(7)           \\
    jQMC (Load-balanced LRDMC) & -8.07870(2) & -17.23472(2) & -25.31459(4)  & -27.6(1.2)  \\
    \Hline
  \end{tabular}
\end{table*}
%%%%%%%%%%%%%%%%%%%%%%%%%%%%%%%%%%%%%

%%%%%%%%%%%%%%%%%%%%%%%%%%%%%%%%%%%%%%%%%%%%%%%%
% Weak-scaling 
%%%%%%%%%%%%%%%%%%%%%%%%%%%%%%%%%%%%%%%%%%%%%%%%

%%%%%%%%%%%%%%%%%%%%%%%%%%%%%%%%%%%%%
% Figure
\begin{figure}[t]
    \centering
    \includegraphics[width=1.0\linewidth]{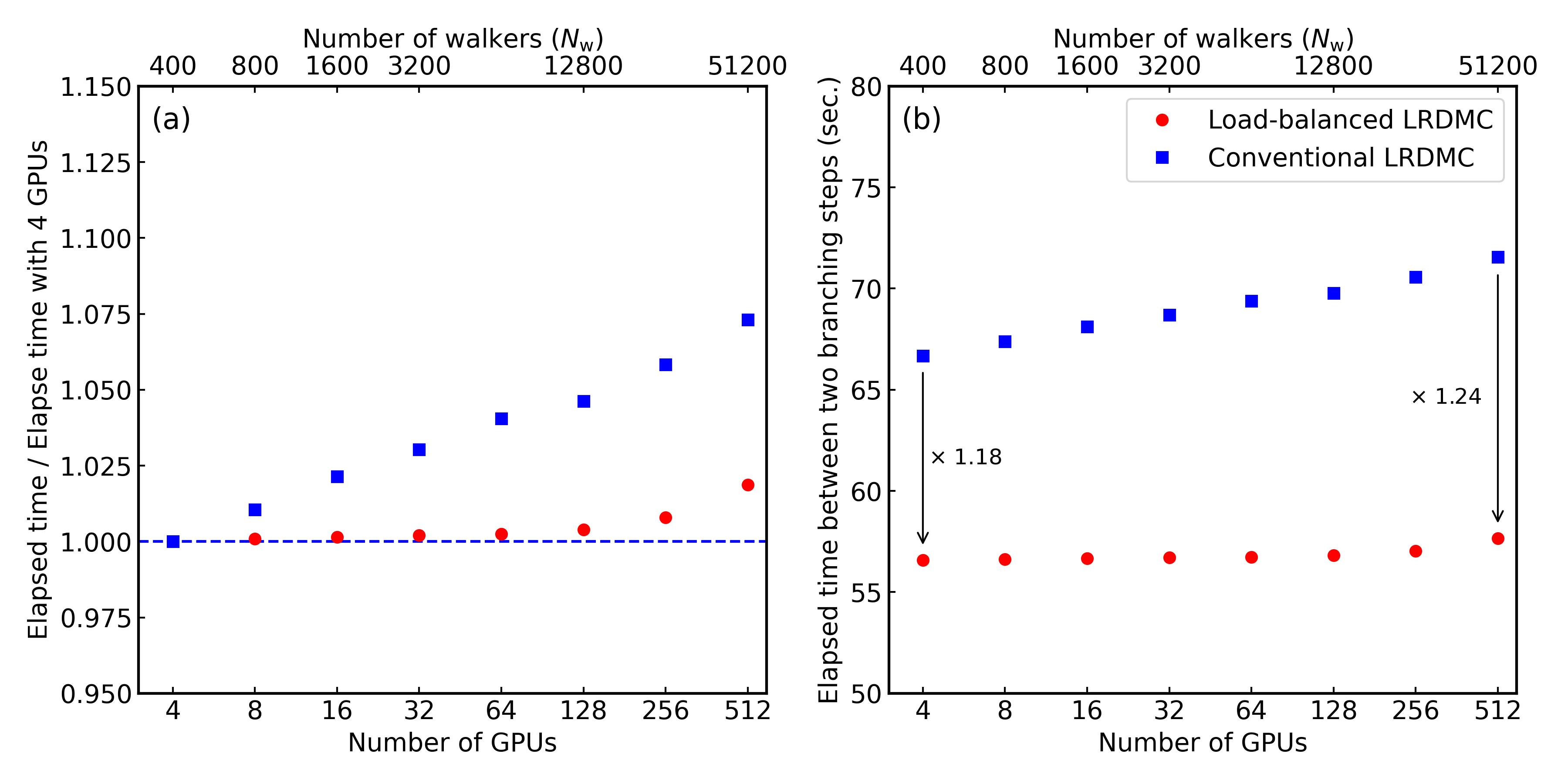}
    \caption[]{Comparison of the weak-scaling benchmark between the conventional and load-balanced LRDMC algorithms, measured on Leonardo using benzene molecule ($N_e$ = 30). A Hamiltonian discretization parameter of $a = 0.30$ a.u. was applied in all LRDMC calculations. In the conventional LRDMC runs, $\tau$ was set to 0.3 a.u., yielding an {\emph{average}} projection number of $N_{\rm proj}=294$. The load-balanced LRDMC calculations employed the same projection number ($N_{\rm proj}=294$). 
    }
    \label{fig:tau-bra-comparison-benchmark}
\end{figure}
%%%%%%%%%%%%%%%%%%%%%%%%%%%%%%%%%%%%%

\section{Weak scaling benchmark 
and direct 
comparison 
between
the conventional and load-balanced LRDMC algorithms}
\label{weak-scaling}
{\vspace{2mm}}
In this study, the weak-scaling benchmark was conducted using the jQMC package on Leonardo~{\cite{2024TUR}}, a supercomputer operated by CINECA in Italy. Leonardo is equipped with four NVIDIA\textsuperscript{\textregistered} A100 GPUs per node. By assigning a large number of walkers to each GPU and utilizing inter-node parallelization, we performed benchmarks for the two different LRDMC algorithms. %Both implementations are those developed within the jQMC package.
For this scaling analysis, we chose the benzene molecule, which has 30 valence electrons ($N_e = 30$). For both carbon and hydrogen atoms, we employed ccECPs pseudopotentials~{\cite{2017BEN, 2018BEN}}, along with the corresponding aug-cc-pVTZ basis sets. For the angular part, the polynominal (cartesian) function was employed (i.e., the $d$ and $f$ orbitals are composed of 6 and 10 orbitals). The trial wavefunction was generated using PySCF~{\cite{2018SUN, 2020SUN}, and subsequently converted to the jQMC wavefunction format via TREX-IO~{\cite{2023POS}}. The Jastrow factor included both two-body and three-body terms~{\cite{2020NAK}}, and was variationally optimized by minimizing the energy using the stochastic reconfiguration (SR) method~{\cite{1998SOR}.
The conventional LRDMC calculations have been performed at an optimal branching time $\tau = 0.30$ a.u., a good compromise between reduction of load-imbalance and increase of stochastic error bars due to walkers fluctuations, yielding an {\emph{average}} projection number of $N_{\rm proj}=294$. The load-balanced LRDMC calculations employed the same projection number ($N_{\rm proj}=294$) to have a clearer idea on the relative efficiency between the two algorithms.

\vspace{2mm}
Fig.~{\ref{fig:tau-bra-comparison-benchmark}}~(a) presents the results of the weak-scaling test for the conventional and load-balanced LRDMC algorithms, each one normalized with respect to the corresponding numerical cost of 400 walkers in 4 GPUs. These benchmarks provide a quantitative assessment of the parallel efficiency of the implementations with multiple walkers and serve as critical indicators of its suitability for large-scale LRDMC simulations. 
As clearly shown in Fig.~{\ref{fig:tau-bra-comparison-benchmark}} (a), the conventional algorithm exhibits a steep decline in computational efficiency as 
a function of the logarithm of the number of walkers (i.e., the degree of GPU parallelization). As explained in the Methods section, this is because the conventional algorithm determines the length of each projection step using a stochastic estimate of the persistence time, resulting in a sizable load imbalance among parallel walkers. Since all walkers must wait until the slowest projection operation is completed, some walkers remain idle in the last projection periods. This behavior leads to an increased likelihood of encountering "slow" walkers with long projection times as the number of walkers grows, resulting in a linear degradation of weak-scaling efficiency as a function of $\ln(N_w)$.
In contrast, the load-balanced LRDMC algorithm ensures, by design, that the computational workload is uniformly distributed among walkers. Consequently, the benchmark results demonstrate that the proposed algorithms maintains nearly optimal weak scaling even for very large number of walkers (up to 51200 in our test).

\vspace{2mm}
Figure~{\ref{fig:tau-bra-comparison-benchmark}}~(b) presents the actual elapsed times between two consecutive branching steps for the conventional and load-balanced LRDMC algorithms. This benchmark demonstrates the extent to which the load-balanced LRDMC algorithm accelerates the calculation compared to the conventional LRDMC algorithm. As clearly shown in Fig.{\ref{fig:tau-bra-comparison-benchmark}}~(b), the load-balanced LRDMC algorithm consistently outperforms the conventional algorithm for any number of $N_w$. The speed-up becomes more pronounced as $N_w$ increases: for $N_w = 400$, the speed-up factor is $\times~1.18$, while for the largest number of walkers tested in this work, $N_w =  51200$, the speed-up reaches $\times~1.24$. It should be noted that the conventional and load-balanced LRDMC calculations were carried out with the same number of branching steps and they provided consistent error bars. These results indicate that the load-balanced LRDMC algorithm not only improves the scaling with respect to $N_w$ compared to the conventional LRDMC algorithm, but also yields a substantial reduction in the actual computational cost.

\section{Conclusions}
\label{conclusion}
In this work, we have introduced a load‐balanced algorithm for the Lattice Regularized Diffusion Monte Carlo (LRDMC) method, which is based on the Green’s Function Monte Carlo (GFMC) framework. We implemented this algorithm in the \emph{ab initio} quantum Monte Carlo packages, TurboRVB and jQMC, and demonstrated that (1) it reproduces the same results as the conventional LRDMC algorithm and (2) it achieves superior parallel efficiency compared to the conventional algorithm as the number of walkers ($N_{\rm w}$) increases up to $\sim$ $10^5$. A direct comparison between the conventional and load-balanced LRDMC calculations reveals that the latter is more efficient by $\times~1.18$ and $\times~1.24$ for $N_{\rm w} = 400$ and  $N_{\rm w} = 51200$, respectively, always due to a more optimal management of the walkers evolution between two branching steps.
Achieving an appropriate load balance among walkers is increasingly critical, as many modern supercomputers are equipped with many-core GPUs. Maximizing GPU utilization requires processing a large number of walkers on each GPU, which in turn necessitates simulations with many walkers. The present work addresses this challenge, enabling more efficient DMC simulations on GPU-based architectures.

%\vspace{2mm}
While our benchmarks were carried out in the context of \emph{ab initio} QMC, the same load‐balanced implementation can be applied wherever GFMC‐based simulations require large walker populations. GFMC is not only widely used in electronic‐structure calculations but also in lattice‐model studies in condensed‐matter physics~{\cite{1998BUO}} and in \emph{ab initio} nuclear‐physics~{\cite{2019LYN}} computations. In any such application where performance is limited by walker imbalance, the load‐balanced LRDMC (GFMC) algorithm presented here should yield significant speedups.

\section{Code and Data availability}
The code and data supporting the findings of this study are available from the jQMC and TurboRVB GitHub repositories:
[\url{https://github.com/kousuke-nakano/jQMC}] and [\url{https://github.com/sissaschool/turborvb}], respectively.
\section*{Acknowledgments}
% computer resource
K.N. and M.C. are grateful for computational resources from EuroHPC for the computational grant EHPC-EXT-2024E01-064 allocated on Leonardo (booster partition). K.N. and M.C. acknowledge EPICURE, a EuroHPC Joint Undertaking initiative for supporting this project on the booster partition of Leonardo  through the EuroHPC JU 2024E01 call for proposals for extreme scale access mode.
% financial support
K.N. acknowledges financial support from the Ministry of Education, Culture, Sports, Science and Technology (MEXT) through Leading Initiative for Excellent Young Researchers (Grant No.~JPMXS0320220025) and from the Japan Science and Technology Agency (JST) through PRESTO (Grant No.~JPMJPR24J9). M. C. thanks the European High Performance Computing Joint Undertaking (JU) for the partial support through the "EU-Japan Alliance in HPC" HANAMI project (Hpc AlliaNce for Applications and supercoMputing Innovation: the Europe - Japan collaboration). 

\vspace{2mm}
% to Sandro
We dedicate this paper to the memory of Prof. Sandro Sorella (SISSA), who passed away in 2023. He was one of the most influential researchers in the QMC community and profoundly inspired this work through his development of the \emph{ab initio} QMC code, TurboRVB. He was the first to propose the load-balanced algorithm described here, and to write its early-version implementation in the TurboRVB package. His former intuition has become even more relevant now, with the advent of python-based codes, jQMC, written in JAX~{\cite{jax2018github}}, which favors computations with many walkers on GPUs, such as the one reported in this work.

\clearpage

%
%
%\usepackage{algorithmic}, \usepackage{algorithm}
%https://qiita.com/jirojiro/items/0ae13aac9112a804f8d5
%%%%%%%%%%%%%%%%%%%%%%%%%%%%%%%%%%
\begin{algorithm*}[t]
    \caption{Conventional LRDMC algorithm.}
\label{alg:LRDMC-tau}

\KwIn{Given $N_{\mathrm{branching}}$, $N_{\mathrm{walker}}$, and $\tau$}
\KwOut{Computed averages and variances}

Generate initial configurations $\vec{x}$ for all walkers\;

\For{$j \le N_{\mathrm{branching}}$}{
  \For{$i \le N_{\mathrm{walker}}$}{
    $w = 1.0$; $\tau_{\rm left} = \tau$ \; 
    \While{$\tau_{{\rm left}} > 0$}{
      Compute the diagonal Green function matrix element $\mathcal{G}_{x',x}$ and $e_L(x)$\;
      Draw a random number, $\xi = [0,1)$ \;
      Compute the persistence time $\tau_{\xi}$, during which the walker stays in configuration $x$:\\
      \hspace{2em} $\tau_{\xi} = \min\left[ \frac{-\log(\xi)}{\sum_{x' \ne x} \mathcal{G}_{x',x}},\; \tau_{\mathrm{left}} \right]$\;
      
      Update the weight: $w \leftarrow w \exp[-\tau_{\xi} e_{L}(x) ]$\;
      
      Update the remaining time: $\tau_{\mathrm{left}} \leftarrow \tau_{\mathrm{left}} - \tau_{\xi}$\;
      
      \If{$\tau_{\mathrm{left}} = 0$}{
        \textbf{break}\;
      }
      
      Move the walker $i$ to a new configuration ($x' \ne x$) with probability:\\
      \hspace{2em} $p_{x',x} = \cfrac{\mathcal{G}_{x',x}}{\sum_{x'' \ne x} \mathcal{G}_{x'',x}}$\;
    }
    Store the accumulated weight $w$ and local energy $e_L$ after the projection\;
    $i \leftarrow i + 1$\;
  }
  Compute the average weight $\bar{w}$ and the average local energy $\bar{e}_L$ before branching\;
  Use branching scheme to control walker weight fluctuations\;
  $j \leftarrow j + 1$\;
}
Compute averages and variances using appropriate estimators\;
\end{algorithm*}
%%%%%%%%%%%%%%%%%%%%%%%%%%%%%%%%%%
%

%%%%%%%%%%%%%%%%%%%%%%%%%%%%%%%%%%%%%%%%%%%%%%%%%%%%%%%%%%%
\begin{algorithm*}[t]
\caption{Load-balanced LRDMC algorithm.}
\label{alg:LRDMC-bra}
\KwIn{Given $N_{\mathrm{branching}}$, $N_{\mathrm{walker}}$, and $N_{\mathrm{proj}}$}
\KwOut{Averages and variances of observables}

Generate initial configurations $\vec{x}_i$ for all walkers\;

\For{$j \gets 1$ \KwTo $N_{\rm branching}$}{
    \For{$i \gets 1$ \KwTo $N_{\rm walker}$}{
        $w = 1.0$ \;
        \For{$k \gets 1$ \KwTo $N_{\mathrm{proj}}$}{
            Compute $b_x$ and $\bar{b}_x$ (diagonal Green function matrix elements)\;
            Update weight: $w \gets w \cdot b_x$, where $b_x = \frac{1}{(\mathcal{W}(x)-E_0)} \cdot \bar{b}_x$\;
            %Propose new configuration $x' \neq x$ with 
            %$Reweight: $w' \gets w \cdot \frac{1}{(\mathcal{W}(x)-E_0)}$\;
            Move walker $x \rightarrow x'$ according to probability $p_{x',x}$\;
        }
        Store the accumulated weight $w$ and local energy $e_L$ after the projection\;
        $i \leftarrow i + 1$\;
    }
  Compute the average weight $\bar{w}$ and the average local energy $\bar{e}_L$ before branching\;
  Use branching scheme to control walker weight fluctuations\;
  Update $E_0$\;
  $j \leftarrow j + 1$\;
}
Compute averages and variances using appropriate estimators\;
\end{algorithm*}
%%%%%%%%%%%%%%%%%%%%%%%%%%%%%%%%%%%%%%%%%%%%%%%%%%%%%%%%%%%

\clearpage

%%%%%%%%%%%%%%%%%%%%%%%%%%%%%%%
% Appendix
%%%%%%%%%%%%%%%%%%%%%%%%%%%%%%%
\appendix

%\section{Appendix}
%\label{appendix}

%\subsection{The stationary distribution of the conventional Lattice regularized Monte Carlo method}
\section{The stationary distribution of the conventional Lattice regularized Monte Carlo method}
\label{app:stationary_distribution_standard}
%\vspace{1mm}
Let the Green’s function with importance sampling be defined as
\begin{equation}
\mathcal{G}_{x',x} = (\Lambda - H_{x',x}) \cdot \frac{\Psi_G(x')}{\Psi_G(x)},
\end{equation}
with $\Lambda \equiv \lambda \delta_{x',x}$ a diagonal matrix.
Since $\mathcal{G}_{x',x}$ is generally not normalized, we introduce a normalization factor $b_x = \sum_{x'} \mathcal{G}_{x',x}$ so that the transition probability can be written as
\begin{equation}
p_{x',x} = \frac{\mathcal{G}_{x',x}}{b_x}.
\end{equation}
Here, we assume that all elements of the matrix $P \equiv p_{x',x}$ are non-negative real numbers after the FN-approximation.
In the GFMC framework, both position $x$ and weight $w$ are updated as follows:
\begin{enumerate}
    \item  Generate $x_{n+1} = x'$ with probability $p_{x',x}$, 
    \item  Update the weight: $w_{n+1} = w_n \cdot b_x$,
\end{enumerate}
implying that we are tracking not only the probability of position, denoted as $\pi(x)$, but also a joint probability distribution over both position and weight, denoted as $\mathcal{P}(x, w)$.
First, we drive the equilibrium probability of position, $\pi(x)$. The master equation for the probability distribution $\pi(x)$ over the state space becomes
\begin{equation}
\pi_{n+1}(x_j) = \sum_{i} \pi_n(x_i) \, p_{x_j, x_i}.
\label{eq:projection-elements}
\end{equation}
Defining the vector
\begin{equation}
\vec{\pi}_n = \left[ \pi_n(x_1), \pi_n(x_2), \ldots, \pi_n(x_M) \right],
\end{equation}
the transition matrix $P \in \mathbb{R}^{M \times M}$ with elements $p_{x',x}$, where $M$ is the number of basis, the above master equation can be expressed compactly as
\begin{equation}
\vec{\pi}_{n+1} = \vec{\pi}_n P.
\label{eq:projection-vector}
\end{equation}
The matrix $P$ is a right stochastic matrix: a square matrix with non-negative entries and rows summing to 1. If $P$ is irreducible (i.e., any state is reachable from any other in a finite number of steps) and aperiodic (i.e., the greatest common divisor of the return times to a state is one), then it is a {\it primitive matrix} and satisfies the conditions of the Perron–Frobenius theorem~{\cite{2023MEYER}}.
By Gershgorin’s circle theorem, the maximum eigenvalue of any right stochastic matrix is 1~{\cite{horn2012matrix}}, and the Perron–Frobenius theorem guarantees that all other eigenvalues have modulus less than 1. Consequently, the iteration of the projections by $P$ filters out the eigenstate with the eigenvalue of 1, and there exists a unique stationary distribution $\vec{\pi}_{\rm eq}$ such that
\begin{equation}
\vec{\pi}_{\rm eq} = \vec{\pi}_{\rm eq} P.
\end{equation}
This describes the mathematical structure behind the power method for iteratively updating probability distributions. In other words, repeatedly applying the update $\pi_{k+1} = \pi_k P$ from any initial distribution leads to convergence toward the stationary state $\pi_{\rm eq}$, as all non-dominant eigenmodes decay exponentially. 
We can immediately show that $\pi_{\rm eq}(x) \propto b_x |\Psi_G(x)|^2$ is indeed a stationary distribution by plugging it into Eq.~{\ref{eq:projection-elements}}:
\begin{eqnarray}
\nonumber \sum_{x} b_x |\Psi_G(x)|^2 \cdot p_{x',x } &=& \sum_{x} |\Psi_G(x)|^2 \cdot \mathcal{G}_{x',x } \\ \nonumber &=& \sum_{x} (\Lambda - H_{x',x}) \Psi_G(x) \Psi_G(x') \\ \nonumber &=& \sum_{x} (\Lambda - H_{x,x'}) \Psi_G(x) \Psi_G(x') \\ \nonumber &=& \sum_{x} (\Lambda - H_{x,x'}) \Psi_G(x) / \Psi_G(x') \cdot |\Psi_G(x')|^2 \\ \nonumber &=&  \sum_{x} \mathcal{G}_{x,x'} |\Psi_G(x')|^2 \\ &=&  b_{x'} |\Psi_G(x')|^2,
\end{eqnarray}
where the hermiticity of the Hamiltonian $ H_{x,x'} = H_{x',x}$ is exploited, by also assuming for simplicity that the Hamiltonian matrix elements are real.
Then, by the Perron–Frobenius theorem, this left eigenvector with eigenvalue 1 is unique, confirming that $b_x |\Psi_G(x)|^2$ corresponds to the stationary distribution (see also Eq.~8.40 of Ref.~{\onlinecite{2017BEC}}).

{\vspace{2mm}}
Next, we derive the equilibrium distribution of position and weight, $\mathcal{P}(x, w)$. Notice that the stationary distribution $\pi(x)$ obtained earlier is the unweighted marginal of $\mathcal{P}(x, w)$, i.e., $\pi(x) = \int dw \, \mathcal{P}(x, w)$. The transition kernel $K$ for the weighted process is defined as
\begin{equation}
K(x', w' | x, w) = p_{x',x} \, \delta(w' - w b_x).
\end{equation}
The corresponding master equation becomes
\begin{equation}
\mathcal{P}_{n+1}(x', w') = \sum_x \int dw \, K(x', w' | x, w) \, \mathcal{P}_n(x, w).
\end{equation}
Multiplying both sides by $w'$ and integrating yields:
%
%\begin{equation}
%\int dw' \, w' \mathcal{P}_{n+1}(x', w') = \sum_x \mathcal{G}_{x',x} \int dw \, w \mathcal{P}_n(x, w)
%= \sum_x p_{x',x} b_x \int dw \, w \mathcal{P}_n(x, w).
%\end{equation}
\begin{equation}
\int dw' \, w' \mathcal{P}_{n+1}(x', w') = 
\sum_x p_{x',x} b_x \int dw \, w \mathcal{P}_n(x, w) =
\sum_x \mathcal{G}_{x',x} \int dw \, w \mathcal{P}_n(x, w).
\end{equation}
The marginal weighted probability $\Pi(x) = \int dw \, w \, \mathcal{P}(x,w)$ is nothing but the many-body wavefunction at step $n$, i.e., $\Psi_n(x) \Psi_G(x)  \equiv \Pi_n(x) = \int dw \, w \, \mathcal{P}_n(x,w)$.
In the vector notation, this leads to:
\begin{equation}
\vec{\Pi}_{n+1} = \vec{\Pi}_n \mathcal{G}.
\end{equation}
Since $\mathcal{G}_{x',x} = p_{x',x} b_x$, we can define a diagonal matrix $B$ with elements $B_{x,x} = b_x$, so that $\mathcal{G} = B P$. If $B$ is strictly positive and $P$ is primitive, then $\mathcal{G}$ is also primitive. Therefore, according to the Perron–Frobenius theorem, the master equation has a stationary (unnormalized) solution $\Pi_{\rm eq}(x)$, and the spectral radius $\rho(\mathcal{G}) > 0$ ensures the existence of a dominant eigenvector. In fact, the eigenvector of $\mathcal{G}$ is $\Psi_0(x) \Psi_G(x)$, and the corresponding eigenvalue is $\Lambda - E_0$, which can be easily verified by substitution.
\begin{eqnarray}
\sum_{x} \Psi_0(x) \Psi_G(x) \cdot \mathcal{G}_{x',x } &=& \sum_{x} \Psi_0(x) \Psi_G(x) (\Lambda - \mathcal{H}_{x',x}) \cfrac{\Psi_G(x')}{\Psi_G(x)}  \\ &=& \sum_{x} \Psi_0(x) (\Lambda - \mathcal{H}_{x',x}) \Psi_G(x') \\ &=& (\Lambda - E_0) \Psi_0(x') \Psi_G(x'). \\
\end{eqnarray}
The eigenvalue ($\Lambda - E_0$) is apparently the dominant eigenvalue because $E_0 < E_i$ for $\forall i$, where $i$ refers to the $i$-th excited states.
Thus, applying $\mathcal{G}$ repeatedly to any initial distribution yields
\begin{equation}
\vec{\Pi}_{\rm init} {\mathcal{G}}^n = \vec{\Pi}_{n} \sim (\Lambda - E_0)^n \Psi_0(x) \Psi_G(x),
\end{equation}
where the prefactor $(\Lambda - E_0)^n$ is canceled out in computing an observable.

%\subsection{The stationary distribution of the load-balanced Lattice regularized Monte Carlo method}
\section{The stationary distribution of the load-balanced Lattice regularized Monte Carlo method}
\label{app:stationary_distribution_load_balance}
%\vspace{1mm}
%Given a fixed-node Hamiltonian $H$ on a lattice, we can formally divide its matrix elements in off-diagonal $\bar{\mathcal{H}}_{x',x}$ and diagonal $\mathcal{W}(x) \equiv \mathcal{H}_{x,x}$ ones:
%
%\begin{equation}
%\mathcal{H}_{x',x} = \bar{\mathcal{H}}_{x',x} + \delta_{x,x^\prime} \mathcal{W}(x)
%\end{equation}
%
%where $\bar{\mathcal{H}}_{x,x}=0$. With the above definition, the Schr\"odinger equation with the fixed-node Hamiltonian can be casted in the following form:
%\begin{equation}
% \sum_{x}\bar{\mathcal{H}}_{x',x} \Psi(x)=\left(E - \mathcal{W}(x') \right) \Psi(x')
%\end{equation} 
Given the definition of the modified Green's function in the load-balanced LRDMC algorithm:
%above relation, with an analogy of the original Schr\"odinger equation, one can define a Green function (with importance sampling) that filters out the ground state:
%
\begin{equation}
\mathcal{G}_{x^\prime,x} = -{1 \over \mathcal{W}(x) - E_0} \bar{\mathcal{H}}_{x',x} \cfrac{\Psi_G(x')}{\Psi_G(x)},
\label{eq:greens_function_load_balance_appendix}
\end{equation}
with $\bar{\mathcal{H}}_{x',x}$ a hermitian (and real for the sake of simplicity) matrix, we can follow the same steps carried out in Appendix \ref{app:stationary_distribution_standard} to show that the marginal weighted probability, which reads:
\begin{equation}
\Pi_{\rm eq}(x) = \int dw \, w \, \mathcal{P}(x,w) = \left( \mathcal{W}(x)-E_0 \right) \Psi_G(x)\Psi_0(x),
\label{eq:marginal_weighted_probability_appendix}
\end{equation}
is stationary in the stochastic Markov chain process driven by $\mathcal{G}_{x^\prime,x}$ in Eq.~\ref{eq:greens_function_load_balance_appendix}, such that
\begin{equation}
\vec{\Pi}_{\rm eq} = \vec{\Pi}_{\rm eq} \mathcal{G}.
\end{equation}
This can be proven as follows:
\begin{eqnarray}
\sum_{x} (\mathcal{W}(x) - E_0) \Psi_0(x) \Psi_G(x) \cdot \mathcal{G}_{x',x } &=& -\sum_{x} \Psi_0(x) \Psi_G(x) {\mathcal{\bar{H}}_{x',x}} \cfrac{\Psi_G(x')}{\Psi_G(x)}  \\ &=& -\sum_{x} \Psi_0(x) {\mathcal{\bar{H}}_{x',x}} \Psi_G(x') \\ &=& (\mathcal{W}(x') - E_0) \Psi_0(x') \Psi_G(x').
\end{eqnarray}
This does not guarantee that the dominant eigenvalue of the master equation is 1, because $\mathcal{G}$ in Eq.~\ref{eq:greens_function_load_balance_appendix} (Eq.~\ref{eq:modified_greens_function} of the main text) is not normalized. However, $\Pi_{\rm eq}$ is certainly a saddle point in the space of distribution functions for those processes driven by $\mathcal{G}$.
Moreover, by performing an eigenvalue decomposition of the initial state $\Psi_\text{init}$:
\begin{equation}
\left| {{\Psi _{\text{init}}}} \right\rangle  = \sum\limits_{i \ge 0} {{a_i}\left| {{\Psi _i}} \right\rangle },
\end{equation}
where ${a_i}$ is the coefficient for the $i$-th eigenvectors (${{\Psi _i}}$) with energy $\epsilon_i$, and $\epsilon_{i+1} \ge \epsilon_i$, one can show that the initial distribution $\Pi_\text{init}(x) = \left( \mathcal{W}(x)- E_0 \right) \Psi_G(x)\Psi_\text{init}(x),$ with $E_0$ the best guess for the ground state energy $\epsilon_0$ (assumed for simplicity non degenerate), is transformed by the application of $\mathcal{G}$ in such a way that the distribution at the next iteration reads:
\begin{equation}
{\Pi}_\text{next}(x) = \left( \mathcal{W}(x)-\epsilon_0 \right) \left [ a_0 \Psi_G(x)\Psi_0(x) + \sum_{i > 0} a_i \frac{\mathcal{W}(x)-\epsilon_i}{\mathcal{W}(x)-\epsilon_0} \Psi_G(x)\Psi_i(x) \right].
\end{equation}
The remaining part beyond $a_0 \Psi_G(x)\Psi_0(x)$ will have renormalized coefficients $a_i \rightarrow a_i \frac{\mathcal{W}(x)-\epsilon_i}{\mathcal{W}(x)-\epsilon_0}$. If $\left|  \frac{\mathcal{W}(x)-\epsilon_i}{\mathcal{W}(x)-\epsilon_0} \right | < 1 ~~ \forall x$, then the amplitude of the higher energy coefficients will be damped and the distribution will converge to $\left(\mathcal{W}(x)-\epsilon_0 \right) \Psi_G(x)\Psi_0(x)$ for a large enough number of applications of $\mathcal{G}$ together with the update $E_0 \rightarrow \epsilon_0$. The condition $\left|  \frac{\mathcal{W}(x)-\epsilon_i}{\mathcal{W}(x)-\epsilon_0} \right | < 1 $ holds for $\epsilon_i < 2 \mathcal{W}(x) - \epsilon_0$. If the initial state is good enough (namely, with higher energy components suppressed, i.e., $a_k \ll 1$ for $k$ such that $\epsilon_k > 2 \mathcal{W}(x) - \epsilon_0$, and with a non-zero overlap with the ground state, i.e., $a_0 \ne 0$), then the distribution will converge towards the mixed ground state one (Eqs.~\ref{eq:marginal_weighted_probability_appendix} and \ref{eq:stationary_load_balance}). The convergence can be easily monitored during the simulation by looking at the decrease (or stability) of $E_0$, estimated during the projection (see Eq.~\ref{eq:ground_state_energy_load_balance}). Indeed, breaking the convergence condition will mean introducing very high energy states, leading to an energy blow-up. In practice, we never found such a case in all systems studied so far, starting from ``regular'' variational trial states. This is related to the fact that $\mathcal{W}(x)$ turns out to be always large compared to the energy scales of the projected wavefunction, yielding $\forall x$ a sufficiently wide energy stability window $[\epsilon_0,2 \mathcal{W}(x) - \epsilon_0]$.

\section{Emulation of the parallelization efficiency of the conventional LRDMC algorithm}
\label{app:emulation}
The degradation of the parallelization efficiency of the conventional LRDMC algorithm can be easily demonstrated using a simple emulation of the implementation (i.e., a toy model). If the total off-diagonal transition probability $b_x = \Sigma_{x' \ne x} \mathcal{G}_{x',x}$ is assumed to follow a normal distribution (or is approximated as a constant), then the number of projection steps required until $\tau_{\rm left} = 0$ can be simulated using a simple random sampling program, corresponding lines 3-5,7-9,11-13,16 of Algorithm~{\ref{alg:LRDMC-tau}}. In Fig.~{\ref{fig:tau-emulation}}, we plot the minimum, maximum, and average number of projection steps required to reach $\tau_{\rm left} = 0$ as a function of the number of walkers used in the emulation. In this emulation, the projection time $\tau$ was set 5.0, and the off-diagonal sum of the Green’s function, $b_x \equiv \sum_{x' \ne x} \mathcal{G}_{x',x}$ in Eq.~{\ref{eq:tau-xi}}, was sampled from a normal distribution with mean $1\times10^3$ and standard deviation 0.5. As the Figure shows, while the average number of steps remains constant regardless of the number of walkers, the branching algorithm must wait for the slowest walker to finish. The outcome of this emulation reproduces the $\log(N_{\rm w})$ scaling of the actual benchmark results shown in Fig.~{\ref{fig:tau-bra-comparison-benchmark}}.

%%%%%%%%%%%%%%%%%%%%%%%%%%%%%%%%%%%%%
% Figure
\begin{figure}
    \centering
    \includegraphics[width=0.7\linewidth]{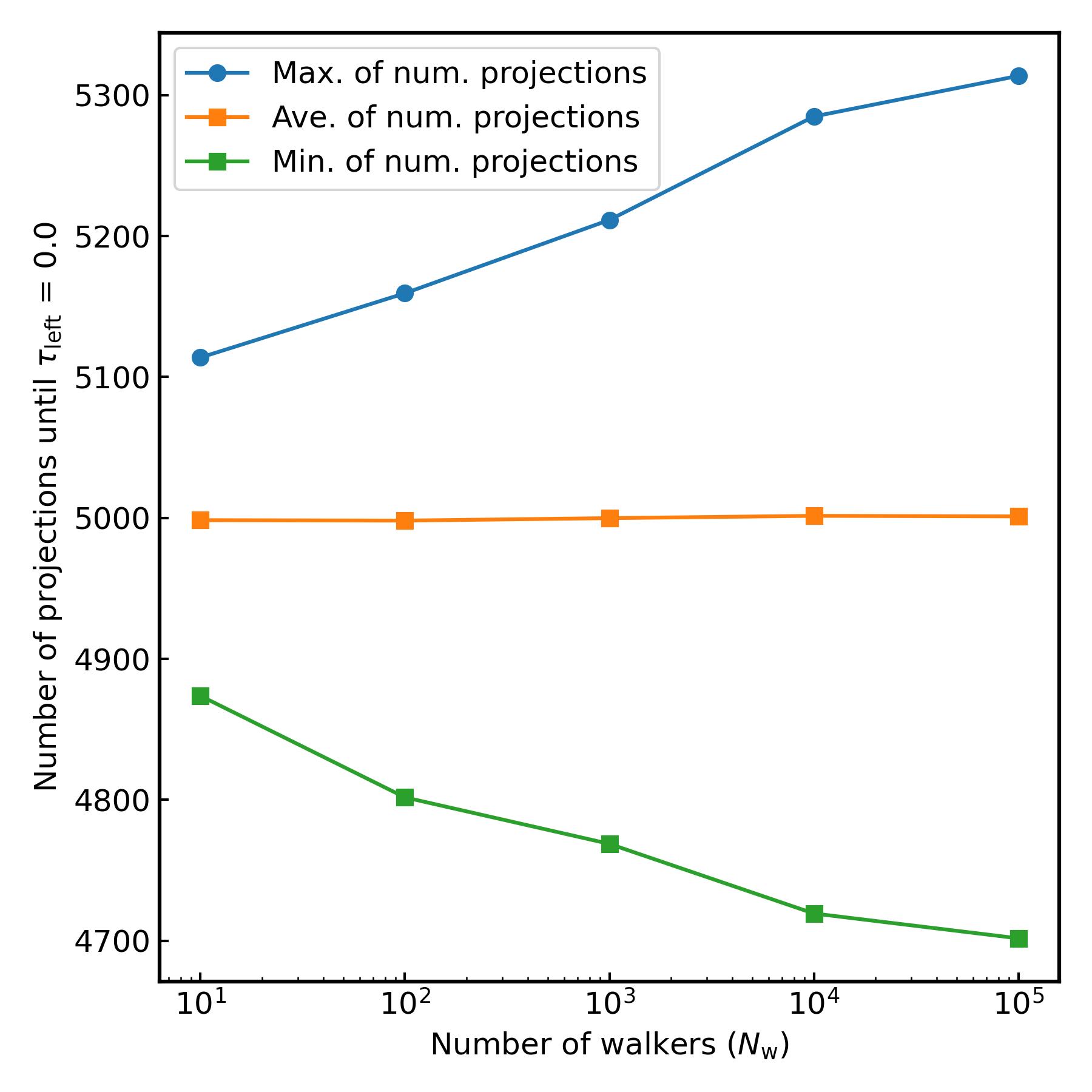}
    \caption[]{The emulation of the evolution of $\tau$ in the conventional LRDMC implementation with respect to the number of walkers (see the main text for the detail). Simulations were carried out for walker populations of $N_{\rm w} = 10,\;100,\;1000,\;10000,$ and $100000$.}
    \label{fig:tau-emulation}
\end{figure}
%%%%%%%%%%%%%%%%%%%%%%%%%%%%%%%%%%%%%

%\subsection{Fixed-node approximation and lattice discretization in LRDMC}
\section{Fixed-node approximation and lattice discretization in LRDMC}
\label{app:fna_and_lattice_regularization}
%\vspace{1mm}
As already mentioned, the Green's function cannot be made strictly positive for fermions; therefore, the fixed-node approximation (FNA) 
has to  be introduced{~\cite{2017BEC}} in order to avoid the sign problem. The FNA is implemented by defining an effective fixed-node (FN) Hamiltonian ${\mathcal{H}}_{x',x}^{\text{FN}}$, with modified off-diagonal matrix elements and with the inclusion of the spin-flip term ${\mathcal{V}_{{\rm{sf}}}}\left( {x} \right) = \sum\limits_{x':{s_{x,x'}} > 0}^{} {{{\mathcal{H}}_{x',x}}{\Psi _\text{G}}\left( {x'} \right)} /{\Psi _\text{G}}\left( {x} \right)$:
\begin{equation}
{\mathcal{H}}_{x',x}^{\text{FN}} = 
 \begin{cases}
  {{\mathcal{H}}_{x,x}} + {\mathcal{V}_{{\rm{SF}}}}\left( {x} \right)\,\,\,\,{\rm{for}}\,\,\,\,x' = x,\\
  {{\mathcal{H}}_{x',x}}\,\,\,\,\,\,\,\,\,\,\,\,\,\,\,\,\,\,\,\,\,\,\,\,\,\,\,\,{\rm{for}}\,\,\,\,x' \ne x,{s_{x',x}} < 0,\\
  0\,\,\,\,\,\,\,\,\,\,\,\,\,\,\,\,\,\,\,\,\,\,\,\,\,\,\,\,\,\,\,\,\,\,\,\,\,\,{\rm{for}}\,\,\,\,x' \ne x,{s_{x',x}} > 0,
\end{cases}
\label{eq:fn-hamiltonian_appendix}
\end{equation}
where ${s_{x',x}} = {\Psi _\text{G}}\left( {x} \right){{\mathcal{H}}_{x',x}}{\Psi _\text{G}}\left( {x'} \right)$. In other words, for off-diagonal matrix elements ${{\mathcal{H}}_{x',x}}$ coming from the Laplacian discretizaton (kinetic terms), ${s_{x',x}} >0$ indicates that the hopping $x \rightarrow x'$ crosses the nodal surface of the guiding function. For more complicated off-diagonal elements, coming e.g. for the non-local pseudopotentials, the sign ${s_{x',x}}$ has a less trivial interpretation, and it is chosen simply to guarantee the non-negativity of the FN ${\mathcal{G}}_{x',x}$. Indeed, the use of the FN Green's function with the so-called guiding function:
\begin{equation}
{\mathcal{G}}_{x',x}^{{\text{FN}}} = \left( {\boldsymbol{\Lambda}  - {\mathcal{H}}_{x',x}^{{\text{FN}}}} \right)\frac{{\Psi _{\text{G}}^{}\left( {x'} \right)}}{{\Psi _{\text{G}}^{}\left( x \right)}} = {\Lambda} \delta_{x', x}  - \frac{{\Psi _{\text{G}}^{}\left( {x'} \right)}}{{\Psi _{\text{G}}^{}\left( x \right)}} {\mathcal{H}}_{x',x}^{{\text{FN}}} 
\end{equation}
can prevent the crossing of regions where the configuration space yields a sign flip of the Green's function; therefore, the walkers are constrained in the same nodal pockets to avoid the sign problem.

The Hamiltonian in the spin flip term is composed of two contributions. ${{{\mathcal{H}}_{x',x}}{\Psi _\text{G}}\left( {x'} \right)} /{\Psi _\text{G}}\left( {x} \right) = {{\mathcal{K}}_{x',x}{\Psi _\text{G}}\left( {x'} \right)} /{\Psi _\text{G}}\left( {x} \right) + {{\mathcal{V}^{\rm nl}_{x',x}}{\Psi _\text{G}}\left( {x'} \right)} /{\Psi _\text{G}}\left( {x} \right)$, where ${\mathcal{K}}_{x,x'}$ and ${\mathcal{V}^{\rm nl}_{x,x'}}$ are the non-local elements in the kinetic and potential terms, respectively. The non-local potential term is considered only for calculations with non-local (semi-local) effective core potentials. With the non-local (semi-local) ECPs, the above fixed-node Hamiltonian corresponds to the T-move~{\cite{2006CAS}} in the standard DMC framework. If one replaces ${{\mathcal{V}^{\rm nl}_{x',x}}{\Psi _\text{G}}\left( {x'} \right)} /{\Psi _\text{G}}\left( {x} \right)$ with ${{\mathcal{V}^{\rm nl}_{x',x}}{D _\text{G}}\left( {x'} \right)} /{D _\text{G}}\left( {x} \right)$, where $D _\text{G}$ represents the determinant part of the trial wavefunction $\Psi _\text{G}$, the resultant fixed-node Hamiltonian corresponds to the T-move with determinant locality approximation (DTM)~{\cite{2019ZEN}} in the standard DMC framework.

% Lattice discretization
In LRDMC, the original continuous Hamiltonian is regularized by allowing electron hoppings with step size $a$, in order to mimic the electronic kinetic energy on the continuum. The corresponding regularized Hamiltonian ${{\hat{\mathcal{H}}}^a}$  is then defined  such that ${{{\hat {\mathcal{H}}}^a}} \to {\hat {\mathcal{H}}}$ for $a \to 0$. Namely, the kinetic and potential parts are approximated by a finite difference form: ${{{\hat {\mathcal{H}}}^a}} \equiv \hat{K}^a + \hat{V}^a$. 

We start from the kinetic part $\hat{K}^a$. The discretized Laplacian $\nabla_a^2$ acts on a function $f\left( {{x_i},{y_i},{z_i}} \right)$ as
\begin{equation}
\nabla^2_{a,i}f\left( {{x_i},{y_i},{z_i}} \right) = \frac{1}{{{a^2}}}\left\{ {\left[ {f\left( {{x_i} + a} \right) - f\left( {{x_i}} \right)} \right] + \left[ {f\left( {{x_i} - a} \right) - f\left( {{x_i}} \right)} \right]} \right\} \leftrightarrow {y_i} \leftrightarrow {z_i}.
\label{eq:nabla_discretization}
\end{equation}
Therefore, the discretized kinetic operator $\hat{K}^a \equiv -\cfrac{1}{2}\sum_i \nabla^{2}_{i,a}$ acts on $\ket{x}$ as:
\begin{eqnarray}
\hat{K}^a \ket{x} \equiv - \cfrac{1}{2} \sum_{i=1}^{N_e} \nabla^2_{a,i} \ket{x} = \cfrac{3N_e}{a^2} \ket{x} - \cfrac{1}{2a^2} \sum_{j=1}^{6N_e} \ket{x_j},
\end{eqnarray}
leading to
\begin{eqnarray}
\hat{K}^a_{x', x} \frac{{\Psi _\text{G}}\left( {x'} \right)}{{\Psi _\text{G}}\left( {x} \right)} = \cfrac{3N_e}{a^2} \delta_{x', x} - \cfrac{1}{2a^2} \frac{{\Psi _\text{G}}\left( {x'} \right)}{{\Psi _\text{G}}\left( {x} \right)} \delta^{ad}_{x', x},
\end{eqnarray}
where $\delta^{ad}_{x', x} =1$ only if $x'$ is adjacent to $x$, otherwise 0. Here, one should consider {\it only} $6N_e$ $x'$ adjacent to $x$. Thus, the GFMC method is applicable to the continuous \emph{ab initio} Hamiltonian, thanks to the sparsity of the discretized kinetic operator. The three versors, upon which the displacements in Eq.~\ref{eq:nabla_discretization} are defined, are set to rotate with random angles along the LRDMC simulation, in such a way that the underlying lattice is \emph{randomized} and the lattice space extrapolation reach the continuous space limit faster.

The potential term $\hat{V}^a$ is divided into the local and non-local term ${\hat{V}^a} \equiv \hat{V}_{\rm loc}^a + \hat{V}_{\rm nl}$, where only the local term depends on the discretized mesh $a$, through the following definition:
\begin{equation}
{V_{\rm loc}^a}\left( {{x}} \right) \equiv V_{\rm loc}\left( {{x}} \right) + \frac{1}{2}\left[ {\frac{{\sum\nolimits_i {\left( {\nabla^2_{a,i} - {\nabla^2_i}} \right) {\Psi _{\text{G}}}\left( {{x}} \right)} }}{{{\Psi _{\text{G}}}\left( {{x}} \right)}}} \right],
\label{eq:lrdmc_potential}
\end{equation}
with $V_\textrm{loc}(x) = \sum_i \nu_{{\rm ei}}(r_i) + \sum_{i,j} \nu_{{\rm ee}}(r_{i,j}) + V_{\rm{ii}}$ electron-ion (ei), electron-electron (ee), and ion-ion (ii) interactions. With the modified potential in Eq.~\ref{eq:lrdmc_potential}, the local energy of the original Hamiltonian coincides with that of the regularized one:
\begin{equation}
e_L^a(x) = \frac{\braket{\Psi_G|-\frac{1}{2}\nabla_a^2 + \hat{V}_{\rm loc}^a + \hat{V}_{\rm nl}|x}}{\braket{\Psi_G|x}} = \frac{\braket{\Psi_G|-\frac{1}{2}\nabla^2 + \hat{V}_{\rm loc} + \hat{V}_{\rm nl}|x}}{\braket{\Psi_G|x}} \equiv e_L(x).
\label{eq:local_energy_condition}
\end{equation}
Together with the randomized mesh, also the condition in Eq.~\ref{eq:local_energy_condition} accelerates the convergence of the FN ${\hat {\mathcal{H}}}^a$ ground-state energy to the continuous $a\rightarrow 0$ limit.

%
%\vspace{2mm}
With the aim at defining a rigorously stable and robust method that works for any finite
value of $a$, we need to further modify the local potential in Eq.~\ref{eq:lrdmc_potential}, to avoid negative divergences that can arise either from the nodes of the guiding function, or from electron configurations too close to the nuclei, or from both, where the bare electron-ion potential is strongly attractive.  
To proceed, we notice that the regularization of Eq.~{\ref{eq:lrdmc_potential}} can be interpreted as the modification of the bare electron-ion potential involving the $i$-th electron $\nu_{\rm ei}(r_i) \rightarrow \nu^{a}_{{\rm zv},i}(x)$, 
where:
\begin{equation}
\nu^{a}_{{\rm zv},i}(x) = \nu_{{\rm ei}}(r_i) + \frac{{\left( {\nabla^2_{a,i} - {\nabla^2_i}} \right) {\Psi _{\text{G}}}\left( {{x}} \right)}}{2{\Psi _{\text{G}}}\left({{x}} \right)} ~~~\textrm{for $i \in [1,N_e]$},
\label{eq:discretized_potential}
\end{equation}
such that $V^a_\textrm{loc}(x) = \sum_i \nu_{{\rm zv},i}(x) + \sum_{i,j} \nu_{{\rm ee}}(r_{i,j}) + V_{\rm{ii}}$.
The above modification cancels out most of the singularities of the attractive potential thanks to the electron-ion cusp conditions fulfilled by $\Psi_G$. However, we can still have unbounded negative values on the nodal surface of the guiding function. A safe possibility to protect the simulation from all these instabilities is given by the following modification of the electron-ion
potential for {\emph{each}} electron $i$:
\begin{equation}
\nu^a_{{\rm{max}},i}(x) = \max[\nu^{a}_{{\rm zv},i}(x), \nu^a_{\rm ei}(r_i)]
\label{eq:single-electron-reg}
\end{equation}
where $\nu^a_{\rm ei}(r_i)$ is defined as:
\begin{equation}
\nu^a_{\rm ei}(r_i) = - \sum_I \cfrac{Z_I}{\max(|r_i - R_I|, a)}.
\end{equation}
for all-electron calculations, which introduces a lower bound $\propto -1/a$, while for calculations with ECPs, $\nu^a_{\rm ei}(r_i) = \nu_{\rm ei}(r_i) = - \sum_I {Z^\text{pseudo}_I} / {|r_i - R_I|}$, because in the latter case the bare Coulomb divergence is perfectly canceled by the localized pseudopotential channels for smooth pseudopotentials, implicitly added to ${\mathcal{V}_{{\rm{sf}}}}$ in Eq.~\ref{eq:fn-hamiltonian_appendix} and coming from the partial localization of $\hat{V}_{\rm nl}$.
Since the condition $\nu^a_{\rm ei}(r_i) > \nu^{a}_{{\rm zv},i}(x)$ is often satisfied for electrons within a lattice-space range from the nuclei, Eq.~{\ref{eq:single-electron-reg}} is applied, in practice, only when the electrons cross the nodal surface with the discretized Laplacian (and the discretized ECP mesh):
\begin{equation}
\nu^{a}_{{\rm opt},i}(x) = 
 \begin{cases}
  \nu^a_{{\rm{max}},i}(x)\,\,\,\,\,\,\,\,\,\,\,\,\,\,\,\,\,\,\,\,\,\,\,\,\,\,\,\,\,\,\,\,\,\,\,{\rm{if}}\,\,\,\, \Psi_{\rm G}(x) \Psi_{\rm G}(x_j) < 0, {\rm{for\,\,at\,\,least\,\,one\,\,}} x_j  \\
  \nu^a_{{\rm{zv}},i}(x)\,\,\,\,\,\,\,\,\,\,\,\,\,\,\,\,\,\,\,\,\,\,\,\,\,\,\,\,\,\,\,\,\,\,\,\,\,\,{\rm{otherwise}},
\end{cases}
\label{eq:final_regularized_potential}
\end{equation}
where $x_j$ indicates the possible configurations proposed by the discretized Laplacian (and ECP).
Within this formulation, and by replacing $\nu^{a}_{{\rm zv},i}(x)$ in Eq.~\ref{eq:discretized_potential} with $\nu^{a}_{{\rm opt},i}(x)$ in Eq.~\ref{eq:final_regularized_potential}, the final regularized potential, $\mathcal{W}(x) = \sum_i \nu^{a}_{{\rm opt},i}(x) + \sum_{i,j} \nu_{{\rm ee}}(r_{i,j})+ {\mathcal{V}_{{\rm{sf}}}}(x) + V_{\rm{ii}}$, is protected from the possible negative energy instabilities close to the nodal surface and/or close to the nuclei, and guarantees the existence of a finite ground-state energy for any $a > 0$.
}

\clearpage

%%%%%%%%%%%%%%%%%%%%%%%%%%%%%%%
\bibliographystyle{apsrev4-1}
\bibliography{./references.bib}
%%%%%%%%%%%%%%%%%%%%%%%%%%%%%%%

\end{document}